%% file: robit.tex
\documentclass[letterpaper,12pt]{article}

\usepackage[linkcolor=blue,urlcolor=blue,citecolor=blue,
colorlinks,linktocpage,bookmarks,breaklinks,pdfstartview=FitH]{hyperref}
\usepackage{natbib}
\usepackage{epsfig}
\usepackage{graphicx}
\usepackage{fancyhdr}
\usepackage[all]{hypcap}
\usepackage{latexsym}
\usepackage{amsmath}
\usepackage{verbatim}
\usepackage{subfig}
\usepackage{setspace}
\usepackage{paralist}
\usepackage{enumerate}

\def\today{\number\day\space\ifcase\month\or January\or February\or March\or
April\or May\or June\or July\or August\or September\or October\or November\or
December\fi\space \number\year}

\allowdisplaybreaks

\usepackage{titlesec}
\titlespacing*{\section}{0pt}{-5pt}{-5pt}
\titlespacing*{\subsection}{0pt}{-5pt}{-5pt}
\begin{document}

\onehalfspacing

\begin{center}

\bf
\Large 

Fully Bayesian Classification with Heavy-tailed Priors for Selection in High-dimensional Features with Grouping Structure

\normalsize
Lai Jiang\footnotemark[3], Longhai Li\footnotemark[1], and Weixin Yao\footnotemark[2]
\end{center}

\onehalfspacing

\bigskip
\centerline{\today}

\input{abstract.tex}

\footnotetext[3]{\onehalfspacing  Lady Davis Institute, Jewish General Hospital,   McGill University, Montreal, QC, H3T1E2, CANADA. Email: \texttt{lai.jiang@mail.mcgill.ca}.}

\footnotetext[1]{\onehalfspacing  Department of Mathematics and Statistics, University of Saskatchewan, Saskatoon, SK, S7N5E6, CANADA. Email: \texttt{longhai@math.usask.ca}. Corresponding author. }

\footnotetext[2]{\onehalfspacing
Department of Statistics, 
University of California at Riverside, 
Riverside, CA, 92521, USA. 
Email: \texttt{weixin.yao@ucr.edu}}

\newpage     

\singlespacing

\input{body.tex}

\section*{Acknowledgements}

The research of Longhai Li is supported by funding from Natural Sciences and
Engineering Research Council of Canada (NSERC), and Canada Foundation of
Innovations (CFI).  The research of Weixin Yao is supported by NSF grant DMS-1461677.

\singlespacing

\bibliographystyle{asa}
\bibliography{tprobit}


\input{appendix.tex}

\end{document}

%% file: abstract.tex
\noindent \textbf{Abstract:} Feature selection is demanded in many modern scientific research problems that use high-dimensional data. A typical example is to find the most useful genes that are related to a certain disease (eg, cancer) from high-dimensional gene expressions.   The expressions of genes have grouping structures, for example, a group of co-regulated genes that have similar biological functions tend to have similar expressions. Many statistical methods have been proposed to take the grouping structure into consideration in feature selection, including group LASSO, supervised group LASSO, and regression on group representatives.  In this paper, we propose a fully Bayesian Robit regression method with heavy-tailed (sparsity) priors (shortened by FBRHT) for selecting features with grouping structure.  The main features of FBRHT include that it discards more aggressively unrelated features than LASSO, and it can make feature selection within groups automatically without a pre-specified  grouping structure.  In this paper, we use simulated and real datasets to demonstrate that the predictive power of the sparse feature subsets selected by FBRHT are comparable with other much larger feature subsets selected by LASSO, group LASSO, supervised group LASSO, penalized logistic regression and random forest, and that the succinct feature subsets selected by FBRHT have significantly better predictive power than the feature subsets of the same size taken from the top features selected by the aforementioned methods. 

%% file: body.tex

\def\mb#1{\mbox{\boldmath $#1$}}

\def\IID {\setlength{\extrarowheight}{-\baselineskip}
	   \, \begin{array}{cc}  
	   \mbox{\tiny IID} \\ 
	   \sim 
	   \end{array} \,
	  \setlength{\extrarowheight}{0pt}
	  }
\def\IG{\mbox{IG}}
\def\G{\mbox{Gamma}}
\def\hblr{\textbf{BPLR} }

\def\mb#1{\mbox{\boldmath $#1$}}
\def\IID{\,\begin{array}{cc} \\[-20pt] \mbox{\tiny IID} \\[-8pt] \sim
\end{array}\,}
\def \given{\,|\,}
\def\mathscript#1{\hbox{\tiny$#1$}}

\def \bxi {\mb x_{i}}
\def \bbeta{\mb \beta}
\def\F21{\,_{2}F_1}
\def\blambda{\mb\lambda}
\def\by{\mb y}
\def\bx{\mb x}
\def\bX{\mb X}
\def\bB{\mb B}
\def\mU{\mathcal{U}}
\def\mS{\mathcal{S}}
\def\tt{$t$}
\def\lp{\mbox{lp}}
\def \tavg {FBRHTavg}
\def \ttop {FBRHTtop}
\def \topt {FBRHTopt}
\def \bayesglm{\texttt{bayesglm}}

\section{Introduction}


The accelerated development of many high-throughput biotechnologies has made it affordable to collect measurements of high-dimensional molecular changes in cells, such as expressions of genes, which are called \textit{features} generally in this paper, and often called \textit{signatures} in life sciences literature. Scientists are interested in selecting features related to a categorical response variable, such as cancer onset or progression.    

Considering the sparsity of important features related to a response, many researchers have proposed to fit classification or regression models with continuous non-convex penalty functions for discovering features related to a response.   It has been widely recognized that non-convex penalties can shrink the coefficients of unrelated  features (noise) more aggressively to 0 than the LASSO while retaining the significantly large coefficients (signal). In other words, non-convex penalties provide a sharper separation of signal from noise.  They are given some new names such as hyper-LASSO or global-local penalties.  Such non-convex penalties include (but not limited to): $t$ with small degree of freedom \citep{gelman2008weakly,yi2012hierarchical}, SCAD \citep{fan2001variable}, horseshoe  \citep{gelman2006prior, carvalho2009handling, carvalho2010horseshoe, polson2012half-cauchy,van_der_pas2014horseshoe},  MCP \citep{zhang2010nearly}, NEG \citep{griffin2011bayesian},   adaptive LASSO \citep{zou2006adaptive}, generalized double-pareto \citep{armagan2010bayesian}, Dirichlet-Laplace and Dirichlet-Gaussian \citep{bhattacharya2012bayesian}; among others.  \citet{kyung2010penalized,polson2010shrink,  polson2012good} and \citet{polson2012local} provide reviews of non-convex penalty functions;  \citet{breheny2011coordinate} and \citet{wang2014optimal} investigate optimization algorithms for learning classification/regression likelihood penalized by non-convex functions, which is often called \textit{non-convex learning} for short.  A non-convex penalty typically corresponds to a prior distribution with heavier tails than Laplace distribution in Bayesian terminology. Hereafter, we will use heavy-tailed priors exchangeably for non-convex penalties.  

Besides sparsity of signal, biological features often have grouping structure or high correlation;  this often has a biological basis, for example  a group of genes relate to the same molecular pathway, or are in close proximity in the genome sequence, or share a similar methylation profile  \citep{clarke2008properties,tolosi2011classification}.  For such datasets, non-convex penalty will make selection within a group of highly correlated features: either splitting important features into different modes of penalized likelihood  or suppressing less important features in favour of more important features.  The within-group selection is indeed a desired property if our goal is selecting a sparse subset of features. Note that the within-group selection does not mean that we will lose other features within a group that are also related to the response because other features can still be identified from the group representatives using the correlation structure.  On the other hand, the within-group selection results in a huge number of modes in the posterior (for example, two groups of 100 features can make $100^{2}$ subsets containing one from each group).   Therefore, optimization algorithms encounter great difficulty in reaching a global or good mode because in non-convex region, the solution paths are discontinuous and erratic. Although superior properties  of non-convex penalties compared to LASSO have been theoretically proved in statistics literature, many researchers and practitioners have been reluctant to embrace these methods due to their lack of convexity, and for good reason: non-convex objective functions are difficult to optimize and often produce unstable solutions~\citep{breheny2011coordinate}. 

There are also other methods that consider directly the grouping structure or the correlation among features in classification/regression models.  The first approach is to fit classification models on the ``new features" constructed from the feature groups (for example centroids, or means), of features within groups; see \citet{jager2002improved, huang2003gene, dettling2004finding, park2007averaged}, \citet{reid2015sparse}, and the references therein. The second approach is to fit classification models with penalties that enforce similarity for the coefficients of features within groups, such as group or fused LASSO \citep{meier2008group, rapaport2008classification}. Group and fused LASSO successfully achieve better predictive performance than plain LASSO, because they consolidate the predictive power of all features within groups. However, since the coefficient of the features within a group are forced to be similar, it becomes harder to select features within groups.  Another problem of this approach is the so-called correlation bias \citep{tolosi2011classification}---the magnitudes of the coefficients of a group will decrease as the group size increases, which results in ambiguity in comparing features across groups when the group sizes vary greatly.  The third approach is the two-stage selection, including  supervised group LASSO (SGL) \citep{ma2007supervised} and ProtoLASSO \citep{reid2015sparse}.  SGL works in two stages: 1) applying LASSO  to the features of each group separately, and then, 2) applying LASSO to the features selected from each group in step 1. ProtoLASSO selects prototype features within each group using marginal correlations, and applies LASSO to the prototype features.  A drawback of SGL and Prototype LASSO is that it makes selection separately in each group, therefore, it cannot consider the joint effects of features from different groups in stage 1. It is likely that a feature is not very useful marginally but becomes very predictive if combined with another feature.  All of the previous three approaches rely on a pre-specified grouping structure, which is often found by a clustering algorithm. However, the functional groups and statistical groups may not match perfectly. In addition, such simple grouping is probably too artificial to explain the complicated biological activities; for example a response may be related to negative correlations among multiple genes. 

 
The fully Bayesian approach---using Markov chain Monte Carlo (MCMC) methods to explore the multi-modal posterior---is a valuable alternative for non-convex learning, because a well-designed MCMC algorithm can travel across many modes for hunting good feature subsets. To the best of our knowledge, the development of fully Bayesian (MCMC) methods for exploring regression posteriors based on heavy-tailed priors has emerged only recently; the relevant articles include \citet{yi2012hierarchical}, \citet{polson2014bayesian}, \cite{zucknick2014mcmc}, 
\citet{piironen2016hyperprior}, \citet{nalenz2017tree},  \citet{johndrow2017scalable}, among others.   In this paper, we develop a sophisticated MCMC method to explore the posterior of Robit model assigned with a class of heavy-tailed priors---Cauchy distribution with small scale. We employ Hamiltonian Monte Carlo method \citep{neal2011mcmc} to draw MCMC samples of the regression coefficients in a restricted Gibbs sampling framework whose computation complexity largely depends on the number of signals rather than the number of all features; this greatly accelerates MCMC sampling for problems with very large dimension.  After MCMC sampling, we divide the samples into subpools according to the posterior modes to find a list of sparse feature subsets. The selection among the list of sparse feature subsets is further aided with cross-validatory evaluation. We will call the above new feature selection method by fully Bayesian Robit Regression with Heavy-tailed Priors, shortened by \textbf{FBRHT}. Compared to other feature selection methods in the literature, FBRHT has the following distinctive characteristics: 
\begin{enumerate} 
\setlength{\itemsep}{-5pt}
 \item FBRHT makes selection within groups automatically \textit{without} a pre-specified grouping structure. Meanwhile, the joint effects of  features from different groups can also be considered. As consequence,  a single feature subset found by FBRHT is more parsimonious than the feature subsets selected by LASSO but retains group representatives. Such succinct feature subsets are much easier to interpret or comprehend based on existing biological knowledge, and easier for further experimental verification or investigation.  
 \item  Because of  within-group selection, the magnitudes of coefficients will not decrease as the number of correlated features increases.
 \item FBRHT extracts a list of parsimonious feature subsets from MCMC samples instead of a single feature subset. Multiple feature subsets  provide multiple explanations of the associations for scientists to further explore.  
 \end{enumerate}  

In this paper, we use simulated datasets with independent and correlated groups of features and two real high-throughput datasets to demonstrate that our MCMC-based non-convex learning method can effectively identify sparse feature subsets with superior out-of-sample predictive performance. Specifically, our empirical results will show that the predictive performances of feature subsets selected by FBRHT are comparable with much larger feature subsets selected by LASSO, group LASSO, supervised group LASSO, random forest,  and non-convex penalized logistic regression. In addition, we will show that the succinct feature subsets selected by FBRHT have significantly better predictive power than the feature subsets of the same size taken from the top features selected by the aforementioned methods. 


This article is organized as follows. In Section \ref{sec:method} we will describe the details of FBRHT in general terms. In Section \ref{sec:sim} we will compare FBRHT with other methods using simulated datasets.  Section \ref{sec:real} reports the results of applying our method to two high-throughput datasets related to Breast Cancer  and Acute Leukemia. This article will be concluded in Section \ref{sec:remark}, with a discussion of relevant future work.

\section{Methodology} \label{sec:method}

\subsection{A Robust Classification Model} \label{sec:T-probitmodel}

Suppose we have collected measurements of $p$ features (such as genes) and a binary response (such as a disease indicator) on $n$ training cases.  For a case with index $i$, we use $y_i$, taking integers $0$ or $1$, to denote the response value  and use a row vector $\mb x_{i}$ to denote the $p$ features, and the first element of $\mb x_{i}$ is set to 1 for including intercept term in linear model.   Throughout this paper, we will use bold-faced letters to denote vectors or matrices. We will write collectively $\by = (y_{1}, \ldots, y_{n})'$, and $\bX=(\bx_{1}', \ldots, \bx_{n}')'$ in which rows stand for observations and columns stand for features. Note that,  we use index $0$ for the intercept term in this paper, ie., the values of the first column of $\bX$ are all equal to $1$, denoted by $\bx_{,0}$.  Using machine learning terminology, we call $(\by,\bX)$ training data, which are used to fit models; in contrast, the data used only in testing the predictive performance is called test data. 

We are interested in modelling the conditional distribution of $y_{i}$ given $\bxi$ for classification and feature selection purpose.  The traditional \textbf{probit} models use a normally distributed auxiliary variable $z_{i}$ to model $y_{i}$ given $\bxi$ as follows:
\begin{equation}
 y_i = I (z_{i}>0), \quad z_i = \bxi  \bbeta + \epsilon_{i}, \quad \epsilon_{i} \sim N(0,1),  \label{eq:probit}
\end{equation}
where $I(\cdot)$ is the indicator function, and $\bbeta$ is a column vector of coefficients with the first element being intercept, denoted by $\beta_{0}$.  With $z_{i}$ integrated out, the above model is equivalent to the following conditional distribution:
$P(y_{i}|\bxi, \bbeta) = \Phi(\bxi  \bbeta)^{y_{i}}(1-\Phi(\bxi  \bbeta))^{1-y_{i}}, \mbox{for } y_{i}=0,1,
$~
where $\Phi$ is the cumulative distribution function (CDF) of the standard normal distribution.  Probit models cannot accommodate some extreme outliers due to the light tails of normals. Typically there exist a large number of extreme outliers in high-throughput data. Therefore, we use a more robust model which replaces the normal distribution for $\epsilon_{i}$ with $t$  distribution:
\begin{equation}
 y_i = I (z_{i}>0) \quad z_i = \bxi\bbeta + \epsilon_i , \quad \epsilon_i \sim  T(\alpha_0, \omega_0),
\end{equation}  
where $T(\alpha, \omega)$ stands for scaled student's $t$ distribution with degrees of freedom $\alpha$, scale parameter $\sqrt{\omega}$, and mean parameter $0$, with a probability density function (PDF) as  
$
t_{\alpha, \omega}(x) = \frac{\Gamma(\frac{\alpha+1}{2}) }{\sqrt{\alpha \pi}\ \Gamma(\frac{\alpha}{2})}  \left(1+\frac{x^2}{\omega \alpha}\right)^{-\frac{\alpha+1}{2}} \frac{1}{\sqrt{\omega}},\  
$
where $\Gamma(\cdot)$ is the Gamma function.  As in probit models, with $z_{i}$ integrated out, the above model is equivalent to the following conditional distribution of $y_{i}$ given $\bxi$:
\begin{equation}
P(y_{i}|\bxi, \bbeta) = T_{\alpha_{0}, \omega_{0}}(\bxi  \bbeta)^{y_{i}}(1-T_{\alpha_{0}, \omega_{0}}(\bxi  \bbeta))^{1-y_{i}}, \mbox{ for } y_{i}=0,1, \label{eqn:model-y}
\end{equation}
where $T_{\alpha, \omega}(x)$ represents the CDF of $T(\alpha, \omega)$, given by
$
T_{\alpha, \omega} ( x) =  \frac{1}{2} + \frac{\Gamma(\frac{\alpha+1}{2}) }{\sqrt{\alpha \pi} \Gamma(\frac{\alpha}{2})} \times  \F21\left(\frac{1}{2},\frac{\alpha+1}{2};\ \frac{3}{2};\ -\frac{x^2}{\alpha\omega}\right) \times \frac{x}{\sqrt{\omega}}, \ 
$
where $\F21$ is the hypergeometric function which is given as the sum of an infinite series \citep{abramowitz1972handbook}.  This model is called Robit model by \citet{liu2004robit}.  It is shown that Robit is more robust to outliers than probit and logistic regression; see \cite{lange1989robust,liu2004robit} and the references therein.  The $\alpha_{0}$ is fixed at $\alpha_{0}=1$, which is appropriate to model the possible range of outliers. In addition, from the CDF of $t$ distribution, we notice that only $\bbeta/\sqrt{\omega_{0}}$ is identifiable in the likelihood of $(\omega_{0}, \bbeta)$ given observation $y_{1}, \ldots, y_{n}$. Therefore, we fix $\omega_{0}$ at some reasonable value. We choose to use $\omega_{0}=0.5$ such that the $T_{\alpha_{0}, \omega_{0}}$ is similar to logistic distribution near origin 0 but has heavier tails than logistic distribution. 

\subsection{Heavy-tailed  Priors (Cauchy) with Small Scales}

In many problems of linking high-dimensional features to a response variable, it is believed that the non-zero regression coefficients are very sparse,  i.e., only very few features are related to the response $y$. In the past decade,  non-convex penalties have drawn attention of many researchers because they can shrink the coefficients of unrelated  features (noise) more aggressively to 0 than the convex $L_{1}$ penalty in LASSO. In other words, non-convex penalties provide a {\em sharper} separation of signal and noise than $L_{1}$.  They are given some new names such as hyper-LASSO, global-local, or selective penalties \citep{polson2012good}.   A non-convex penalty often corresponds to a prior distribution with tails heavier than Laplace distribution (which corresponds to $L_{1}$) in Bayesian methodologies.   In Bayesian interpretation, a typical sample of $\bbeta$ from a heavy-tailed prior has {\em a few} extraordinarily large values representing {\em related features} and many small ones representing {\em unrelated features}.  Therefore, heavy-tailed priors express more closely our belief about $\bbeta$ than Laplace prior.  Among many non-convex penalties, penalty functions corresponding to prior distributions with the same tail heaviness as Cauchy have been shown to have superior performance than $L_{1}$ in detecting very sparse signals; the remarkable such penalties include  horseshoe  \citep{carvalho2009handling, carvalho2010horseshoe, polson2010shrink,  polson2012good, polson2012local, polson2012half-cauchy},  and normal-exponential-gamma (NEG) \citep{griffin2011bayesian}.  Horseshoe and NEG priors have the same heaviness in tails (converging to 0 in the rate of $1/\beta^{2}$) as Cauchy, but they have non-differentiable log PDF at $0$, therefore,  small signals can be shrunken to exact $0$ in penalized likelihood. These two modified priors have been compared with plain Cauchy, and we have found that they produce almost the same results as given by Cauchy. 
The predictive performance of classification models using $t$ priors with various tail heaviness (degrees of freedom) have also been compared empirically on synthetic datasets with very sparse signal, and we have found that Cauchy appeared to be optimal; 
this result confirm the success of penalized likelihood methods using horseshoe and NEG penalties.  On the other hand, using these modified priors demand additional computation in sampling the hyperparameters (local variances for each $\beta_{j}$, i.e., $\lambda_{j}$ below). The additional computation indeed accounts for half of the whole sampling time after we use a restricted Gibbs sampling scheme to greatly shorten the sampling time for regression coefficients.  Therefore,  we chose to use plain Cauchy prior in this article, i.e., $t$ with degree of freedom $\alpha_1 = 1$, denoted by $\beta_j \sim T(\alpha_1, \omega_1),   \mbox{for } j = 1,\ldots, p.$ For MCMC sampling consideration, we express $t$ prior for $\bbeta$ as a scale-mixture normal by introducing a {\em latent} variance $\lambda_{j}$ for each $\beta_{j}$, which is written as follows:
\begin{eqnarray}
\beta_j | \lambda_j  &\sim&  N(0, \lambda_j),  \label{eqn:prior-beta} \\
\lambda_j  &\sim&  \text{Inverse-Gamma}\left(\frac{\alpha_1}{2},\frac{\alpha_1 \omega_1}{2 }\right). \label{eqn:prior-lambda}
\end{eqnarray}
Hereafter, we will write collectively $\blambda = (\lambda_{1}, \ldots, \lambda_{p})$. 

In order to shrink small coefficients toward 0, we need to choose a very small scale parameter $\sqrt{\omega_{1}}$ for Cauchy.  A typical method in Bayesian methodologies for avoiding assigning a fixed value to a parameter is to treat it as a hyperparameter such that it will chosen automatically during MCMC sampling according to marginalized likelihood. However, we have found that this approach does not choose sufficiently small scale to yield very sparse $\bbeta$ because a classification model with $p$ features can easily overfit a dataset with sample size $n\ll p$.  For enforcing sparsity in $\bbeta$ and for reducing difficulty in MCMC sampling, we choose to fix $\sqrt{\omega_{1}}$ at a small value $e^{-5}\approx 0.01$.   A number of upper-tailed quantiles of $|\beta_{j}|$ where $\beta_{j}\sim \mbox{Cauchy}(0, e^{-5})$  are shown in Table~\ref{tab:cauchyprior}. 
\begin{table}[ht]
\begin{center}
\caption{Upper-tailed quantiles of absolute Cauchy with scale $e^{-5}$. }\label{tab:cauchyprior}

\begin{tabular}{lrrrrrrr}
  \hline
Upper probability & 0.200 & 0.100 & 0.020 & 0.010 & 0.002 & 0.001 & 0.0001\\ 
  Quantile of $|\beta_{j}|$ & 0.022 & 0.044 & 0.223 & 0.446 & 2.228 & 4.456 & 42.895\\ 
   \hline
\end{tabular}
\end{center}
\end{table}
We see that this prior postulates that 2 out of 1000 features have coefficients with magnitude $\geq 2.228$, i.e., are related to the response. We believe that this is an appropriate level of sparsity in many problems using high-dimensional features. Another important reason that we can  fix $\sqrt{\omega_{1}}$ is the  ``flatness'' (heaviness) in Cauchy  tails. Due to the ``flatness'' in tails,  very small shrinkage is applied to large coefficients. Since the shrinkage is small,  the estimates of large coefficients are robust to $\sqrt{\omega_{1}}$; see demonstrations in \citet[Fig. 3]{carvalho2010horseshoe}
. This is a distinctive property of  priors with tails as heavy as Cauchy compared to Gaussian, Laplace, and other priors with similar heaviness in tails, for which a careful choice of scale must be made because the shrinkage of large coefficients are large and sensitive to the scale. Therefore, although $\sqrt{\omega_{1}}$ is fixed at a very small value like $0.01$, the prior does not over-shrink large signal, and can accommodate a wide range of signal. 

\subsection{Restricted Gibbs Sampling with Hamiltonian Monte Carlo}\label{sec:mcmc-T-probit}

There is  great difficulty in maximizing the penalized likelihood function using heavy-tailed and small-scaled priors.  For example, using  a small scale for $\sqrt{\omega_{1}}$ such as $e^{-5}$, the R function \texttt{bayesglm} in R package \texttt{ARM} (which implements penalized logistic regression with Cauchy priors) will converge to a mode where almost all coefficients are shrunken to very small values, even when the number  of features ($p$) is small.  On the other hand, using the default 2.5 value, \texttt{bayesglm} does not provide a sparse solution (to be presented in this article). The difficulty in optimization is further intensified by the severe multi-modality in the posterior because heavy-tailed and small-scaled priors can split coefficients of a group of correlated features into different modes rather than shrinking them simultaneously as Gaussian priors do.  Therefore,  although good theoretical properties of non-convex penalties have been proved in statistics literature \citep[e.g.][]{zhang2010nearly}, many researchers and practitioners have been reluctant to embrace these methods because optimization algorithms often produce unstable solutions~\citep{breheny2011coordinate}. This motivated us to develop MCMC algorithms for exploring the posterior with many modes due to the use of heavy-tailed priors. 

Our MCMC algorithm will sample from the joint posterior, $f(\bbeta,\blambda|\by,\bX)$, which is based on the hierarchical models given by equations~\eqref{eqn:model-y},\eqref{eqn:prior-beta}, \eqref{eqn:prior-lambda} with $\alpha_{0},\omega_{0}, \alpha_{1},\omega_{1}$ fixed (so omitted in the following model descriptions). The log posterior can be written as follows:
\begin{equation}
\log (f(\bbeta,\blambda|\by,\bX) ) =\sum_{i=1}^{n} \log (P(y_{i}|\bxi, \bbeta)) + \sum_{j=0}^{p}\log (f(\beta_{j}|\lambda_{j})) + \sum_{j=1}^{p}\log (f(\lambda_{j})) + C, \label{eqn:log-jointpost}
\end{equation}
where the first three terms come from the models defined by \eqref{eqn:model-y},\eqref{eqn:prior-beta}, \eqref{eqn:prior-lambda} respectively,  and $C$ is the log of the normalization constant unrelated to $\bbeta$ and $\blambda$.  The first three terms in \eqref{eqn:log-jointpost} are given as follows:
\begin{eqnarray}
  \log (P(y_{i}|\bxi, \bbeta) ) &=& y_{i}\log(  T_{\alpha_0, \omega_0} (\bxi \bbeta)) + (1-y_{i}) \log( T_{\alpha_0, \omega_0} (-\bxi\bbeta)) \equiv \lp(y_{i}|\bxi\bbeta) , 
 \label{eqn: logprob-y} \\
 \log(f(\beta_{j}|\lambda_{j})) &=& -\frac{1}{2}\log (\lambda_{j}) - \frac{\beta_{j}^{2}}{2\lambda_{j}} + C_{1},  \mbox{  for } j = 0,\ldots, p
 \label{eqn:logprior-beta}\\
 \log(f(\lambda_{j})) &=& -\left(\frac{\alpha_{1}}{2} + 1 \right)\log (\lambda_{j}) - \frac{\alpha_{1}\omega_{1}}{2\lambda_{j}} + C_{2},  \mbox{  for } j = 1,\ldots, p.
 \label{eqn:logprior-lambda}
\end{eqnarray}
where $C_{1}, C_{2}$ are two constants unrelated to $(\bbeta,\blambda)$; the function $\lp(y_{i}|\bxi\bbeta)$ is introduced to indicate that the probability of $y_{i}$ given $\bxi$ is a function of $\bxi\bbeta$.  An ordinary Gibbs sampling procedure to draw samples from \eqref{eqn:log-jointpost} is to alternatively draw samples from the conditional posterior of $\blambda$ given $\bbeta$ with a log density equal to the sum of the last two terms of \eqref{eqn:log-jointpost}, and draw samples from the conditional posterior of $\bbeta$ given $\blambda$ with a log density equal to the sum of the first two terms of \eqref{eqn:log-jointpost}.

The challenge in sampling from the \eqref{eqn:log-jointpost} comes from two aspects of high-dimensional features. One is the high dimension $p$ of $\bbeta$ (or $\bX$); the other is the high correlation among features $\bX$, which results in the high correlation in the conditional posterior of $\bbeta$ given $\blambda$, and correspondingly the multi-modality in the marginal posterior of $\bbeta$ (with $\blambda$ integrated out). To combat these two difficulties, we propose an MCMC sampling algorithm that uses Gibbs sampling with Hamiltonian Monte Carlo (HMC) for sampling $\bbeta$ in a restricted way.  Our MCMC algorithm is sketched below and followed with explanations:

{\sf
\noindent Starting from a previous state for $(\bbeta, \blambda)$,  a new state denoted by $(\hat{\bbeta}, \hat{\blambda})$ is obtained with these steps

\begin{enumerate}[{\em Step} 1:]
\setlength{\itemsep}{0pt}

\item For each $j$, draw a new $\hat{\lambda}_{j}$ from the conditional distribution $f(\lambda_{j}|\beta_{j})$ with log PDF  equal to the sum of  \eqref{eqn:logprior-beta} and \eqref{eqn:logprior-lambda}. It is well-known that $\lambda_{j}$ given $\beta_{j}$ has an Inverse-Gamma distribution given as follows:   
$
\lambda_j | \beta_j \sim \mbox{Inverse-Gamma}\left(\frac{\alpha_1 + 1}{2}, \frac{\alpha_{1}\omega_{1}+\beta_j^2}{2}\right). \label{eqn:post-lambdaj}
$

\item With the new values of $\hat{\lambda}_j$ drawn in step 1, determine a subset, $\bbeta_{U}$, of $\bbeta$ to update in step 3 below. We update $\beta_j$  if $\hat{\lambda}_j$ is large enough. That is, given a pre-scribed threshold value $\eta$, the subset is defined as $U = \{ j | \hat{\lambda}_j > \eta \}$.  The $\bbeta_{U}$ is defined as $\{\beta_{j}|j \in U\}. $ The subset of $\bbeta_{F}=\{\beta_{j}|j \in F = \{0,\ldots, p\} \setminus U\}$ will be kept unchanged in step 3. 

\item Update the set of $\beta_{j}$ with $j \in U$, denoted by $\bbeta_{U}$, by applying HMC to the conditional distribution of $\bbeta_{U}$ given as follows:
\begin{eqnarray}
\lefteqn{\log(f(\bbeta_{U}|\bbeta_{F}, \hat{\blambda}, \bX, \by))} \nonumber \\
&=& \sum_{i=1}^{n} \lp(y_{i}| \mb x_{i,U} \bbeta_{U} + \mb x_{i,F} \bbeta_{F}) +\sum_{j \in U}\log (f(\beta_{j}|\hat{\lambda}_{j})) + C_{3},  \label{eqn:logpost-betaU}
\end{eqnarray} 
where the function lp for computing log likelihood is defined in \eqref{eqn: logprob-y}, and $\mb x_{i,U}$ is the subset of $\bxi$ with feature index in $U$.   After updating $\hat \bbeta_{U}$, the new value of $\bbeta$  is denoted by $\hat{\bbeta}$ in which $\bbeta_{F}$ does not change. Note that, because HMC is a Metropolis algorithm, the new $\hat{\bbeta}$ may be equal to $\bbeta$ if a rejection occurs. 

\item Set $(\bbeta,\blambda)=(\hat\bbeta,\hat\blambda)$, and go back to step 1 for the next iteration. 
\end{enumerate}

}

A typical sampling method for classification models is to augment a latent continuous value $z_{i}$ for each  categorial variable $y_{i}$ \citep{holmes2006bayesian}, and sample from the joint distribution of $z_{1:n}$ along with $\bbeta$ and $\blambda$ \citep[see e.g.][]{zucknick2014mcmc} with Gibbs sampling;  we then can borrow algorithms developed for regression models with heavy-tailed priors \citep{polson2014bayesian, piironen2016hyperprior, nalenz2017tree, johndrow2017scalable}.  Given $\lambda_{j}$, the prior for $\beta_{j}$ is a normal distribution. It is well-known that the  posterior of $\bbeta$ for normal regression given normal priors is a multivariate normal distribution with a covariance matrix involving $\bX'\bX$. Note that this multivariate normal has a dimension $p$.  When $p$ is very large (e.g. thousands), drawing independent samples from a  multivariate normal is extremely inefficient, because the required computation time for decomposing the covariance matrix will increase in the order of $p^{3}$.  Therefore, for drawing samples from $f(\bbeta|\blambda, \bX, \by)$, we choose to use Hamiltonian Monte Carlo (HMC), a special case of Metropolis-Hasting (M-H) algorithms, which explore the posterior in a local fashion without the need to decompose a high-dimensional matrix.  HMC requires computing the log-posterior and its gradient. The gradient of $\log(f(\bbeta|\blambda, \bX, \by))$ given by the following expression:
\begin{equation}
\begin{aligned}
\frac{\partial \mathcal{U}}{\partial \beta_j} &= 
\sum_{i=1}^n \left[ 
 {\Gamma\left(\frac{\alpha_0+1}{2}\right) \over \sqrt{\alpha_0\pi}\ \Gamma\left(\frac{\alpha_0}{2}\right) }\times
 {x_{ij} \over \sqrt{\omega_{0}} }\times
 \dfrac{ 
 \left(1+\dfrac{(\mb x_{i, U}\bbeta_{U}+\mb x_{i, F}\bbeta_{F})^2}{\alpha_0\omega_0}\right)^{-\frac{\alpha_0+1}{2}}
  }
  {1-y_{i} - T_{\alpha_0,\omega_0}(\mb x_{i, U}\bbeta_{U}+\mb x_{i,F}\bbeta_{F}) } \right]+
              \frac{\beta_j}{\hat \lambda_j} ,
\end{aligned}
\end{equation}
where  $\mathcal{U}$ is the function defined in \eqref{eqn:logpost-betaU}.  We can see that once the linear combination $\bX\bbeta$ has been computed,  the log posterior and its gradient can be obtained with very little computation. Computing $\bX\bbeta$ is significantly cheaper than decomposing a matrix of dimension $p$. However, the random-walk behaviour of ordinary M-H algorithms limits the sampling efficiency of M-H algorithms.  In HMC,  the gradient of log posterior is used  to construct a trajectory along the least constraint direction, therefore, the end point of the trajectory is distant from the starting point, but has high probability to be accepted;  for more discussions of  HMC, one is referred to a review paper by \citep{neal2011mcmc}.  


From the above discussion, we see that obtaining the value of $\bX\bbeta$ is the primary computation in implementing HMC. To further accelerate the computation for very large $p$, we introduce a trick called restricted Gibbs sampling; this is inspired by the fact that  updating the coefficients with small $\lambda_{j}$ (small prior variance in the conditional posterior of $\beta_j$ given $\blambda$) in HMC does not change the likelihood as much as updating the coefficients with large $\lambda_{j}$ but updating $\beta_j$ with small or large $\lambda_j$  consumes the same time. Therefore, we use $\hat \blambda$ in step 2 to select only a subset of $\bbeta$, denoted by $\bbeta_{U}$, those have large prior variance $\lambda_{j}$, to update in step 3 (HMC updating). We can save a great deal of time for computing $\bX\bbeta$ in step 3 by caching values of $\bX_F \bbeta_{F}$ from the previous iteration because it does not change in the whole step 3; this greatly accelerates the construction of HMC trajectory.  We typically choose $\eta$ in step 2 so that only $10\%$ of $\bbeta$ are updated in step 3.  

We clarify that  although $\bbeta_F$ (sometimes the whole $\bbeta$) are kept the same in an iteration, the choice of $U$ in step 2 for the next iteration will be updated because $\blambda$ will be updated in step 1.  Thus,  $\beta_j$ will not get stuck to a very small absolute value, unlike that in optimization algorithms this typically occurs. 

The above restricted Gibbs sampling is a valid Markov chain transition for the joint posterior \eqref{eqn:log-jointpost}.  To understand this, let us recall that, in Gibbs sampling we can arbitrarily choose any variables to update with a Markov chain transition that leaves the conditional distribution of chosen variables invariant, provided that the choice of variables to be updated does not depend on the {\em values} of the chosen variables in the previous iteration.    For example, it is not a valid Markov chain transition if we choose $\beta_j$ with large $|\beta_j|$ in the previous iterations; by contrast, it is a valid Markov chain transition if we choose $\beta_j$ to update  by referring to variances of $\bbeta$. In step 3, the choice of $\bbeta_U$ does not depend on the values of $\bbeta$ in the previous step. Instead, the choice only depends on the value of $\lambda_j$ in the previous step, which partially determines the variances of $\bbeta$ in $f(\bbeta|\hat \lambda, \bX, \by)$. Therefore, the updates of $\bbeta_U$ in step 3 is reversible with respect to    $f(\bbeta|\hat \lambda, \bX, \by)$. 


The advantage of HMC is that it can explore highly correlated posterior quickly with a long leapfrog trajectory without suffering from the random-walk problem. This ability of HMC also plays an important role in travelling quickly between multiple modes of the posterior. This is explained as follows. When $\hat\lambda_{j}$ and $\hat\lambda_{k}$ for two correlated features $j$ and $k$ are large after a draw in step 1, the joint conditional posterior of $(\beta_{j}, \beta_{k})$ given $(\hat\lambda_{j}, \hat\lambda_{k})$ are highly negatively-correlated. For such distributions, HMC can move more quickly than random-walk algorithms along the least constrained direction, and this move will lead to the change of modes in joint distribution of $(\beta_{j}, \beta_{k})$ with $\blambda$ integrated out .

There are a huge number of modes in the posterior even when $p$ is moderate when there are a large number of correlated features.  In the empirical studies reported in this paper, we use a two-stage procedure. In \textbf{Stage 1}, we run the restricted Gibbs sampling with HMC using the dataset containing all $p$ features. Then we calculate MCMC means of all coefficients $\bbeta$ and choose only the top $p^{*}=100$ features with largest absolute values of MCMC means. The stage 1 is very time consuming. In \textbf{Stage 2} we re-run the MCMC sampling with only the selected features once again. Our feature selection will be based on the MCMC samples obtained from Stage 2.  A list of setting parameters with recommended values for ease in reference are given in Section~\ref{sec:setting-mcmc}. 

\subsection{Extracting Feature Subsets from MCMC Samples} \label{sec:mcmc-divide}

We run the MCMC sampling as described in Section \ref{sec:mcmc-T-probit} to obtain samples from the posterior of $\bbeta$. With the intercept $\beta_{0}$ removed, this sample is denoted by a matrix $\bB=(\beta_{j,i})_{p\times R}$, in which $\beta_{j,i}$ represents the value of $\beta_j$ in the $i$th sample and $R$ is the number of MCMC samples. The posteriors of $\bbeta$ for Robit models with heavy-tailed priors are severely multi-modal.  For a demonstration, one can look at Figure \ref{fig:toy-mcmc}, which shows a scatterplot of MCMC samples of two $\beta_{j}$'s for two correlated features.   Therefore, we should divide the whole Markov Chain samples $\bB$ into subpools according to the mode that each sample represent.  However, the number of such feature subsets may be huge even the number of features $p$ is mall. Therefore, we only consider dividing Markov Chain samples according to the multiple modes for the Markov Chain samples obtained in {\bf Stage 2} in which a large number of weakly related features have been omitted.  In this article, we use a scheme that looks at the relative magnitude of $\beta_{j}$ to the largest value in all features.  The scheme is rather ad-hoc. However, it is very fast and works well in problems of moderate dimension, such as $p=100$. More advanced methods for extracting feature subsets from MCMC is our priority for future research; see more details in Section~\ref{sec:remark}. The scheme used in this article is described as follows:
\begin{enumerate}[{\em Step} 1:]
\setlength{\itemsep}{0pt}
\item We set $I_{j,i}=1$ if $|\beta_{j,i}| > 0.1 \times \max \{|\beta_{1,i}|, \ldots, |\beta_{p,i}|\}$, and $I_{j,i}=0$ otherwise. By this way, we obtain a boolean matrix $(I_{j,i})_{p\times R}$ with its entry $I_{j,i}$ denotes whether the $j$th feature is selected or not in $i$th sample. 

\item Discard the features with overall low frequency in step 1.   We calculate $f_{j}=\frac{1}{R} \sum\limits_{i=1}^R I_{j,i}$. We will discard a feature $j$ if $f_{j}$ is smaller than a pre-defined threshold, which is set to be $5 \%$ as an ad-hoc choice in this article.  Let $D=\{j|f_{j} < 5\%\}$.  For each $j\in D$, we set $I_{j,i} = 0$ for all $i=1,...,R$. This step is to remove the features that come into selection in step 1 due to MCMC randomness. 

\item Find a list of feature subset by looking at the column vectors of $I$. Each unique column in $I$ represents a different feature subset. 
\end{enumerate}

\subsection{Cross-validatory Predictive Evaluation of Feature Subsets}\label{sec:3.7} \label{sec:fsubset-cv}

The frequencies of feature subsets in MCMC samples may not exactly reflect the predictive power of feature subsets found with the MCMC dividing algorithm presented in Section \ref{sec:mcmc-divide}. Since the feature subsets found by FBRHT are very sparse, it is light in computation to evaluate the predictive power of each feature subset using leave-one-out cross-validation (LOOCV) using the same training cases for simulating MCMC. 

Suppose we want to evaluate the predictive power of a feature subset $\mS$. Alternately for each $i=1,\ldots,n$, we remove the $i$th observations, obtaining a dataset containing only the features in $\mS$ and other observations, denoted by $(\by_{-i},\bX_{-i,\mS})$; we apply \texttt{bayesglm} function (in the R add-on package \texttt{arm}) with default settings  to fit the logistic regression model with $t$ penalty \citep{gelman2008weakly} to $(\by_{-i},\bX_{-i,\mS})$, obtaining an estimate $\hat\bbeta$; the predictive probability for the removed case $i$: $P(y_{i}=c|\bx_{i}, \bbeta)$, for $c = 0,1$, is estimated by plugging the $\hat\bbeta$, that is, $\hat P_{i}(c)\equiv P(y_{i}=c|\bx_{i}, \hat\bbeta)$, for $i=1,\ldots,n, c = 0, 1$.

We will then evaluate the goodness of $\hat P_{i}(c)$ for $i=1,\ldots,n, c = 0,1$ with the actually observed class labels $y_{1},\ldots, y_{n}$.  There are a few criteria available to measure the predictive power.  The first criterion is \textbf{error rate}. We predict  $y_{i}$ by thresholding $\hat P_{i}(1)$ at $0.5$, that is, $\hat y_{i}=1$ if $\hat P_{i}(1)> 0.5$, otherwise, $\hat y_{i}=0$. The error rate is defined as the proportion of wrongly predicted cases: $\mbox{ER}=\frac{1}{n}\,\sum_{i=1}^n\,I(\hat{y}_{i}\not=y_{i})$.  This criterion is very useful to evaluate the performance of the predictions at the boundary $0.5$, but does not punish the very small predictive probabilities at the true labels.  The second criterion is defined as the average of minus log predictive probabilities (\textbf{AMLP}) at the actually observed $y_{i}$: $\frac{1}{n}\,\sum_{i=1}^n\,-\log(\hat P_{i}(y_{i}))$.  AMLP punishes heavily the small predictive probabilities at the true class labels. However, AMLP is very sensitive to only 1 case with poor predictive probability.  The third method is the area under ROC (receiver operating characteristic) curve (\textbf{AUC}).  The ROC curve is the line linking the true and false positive rates as a function of the thresholding $u\in (0,1)$. The AUC is the area under the ROC curve. A larger AUC indicates a better set of predictive probabilities. AUC criterion is less sensitive to a few cases with poor predictions. We use an R-package called \texttt{pROC} \citep{robin2011proc:} to compute AUC.  


After processing  MCMC samples of $\bbeta$ as above, we obtain a list of feature subsets with a column  ``freqs'' indicating the frequencies that each feature subset appears in MCMC samples, and three columns ``cvER'', ``cvAMLP'', and ``cvAUC'',  indicating the LOOCV predictive goodness. 


\subsection{Out-of-sample Predictions}\label{sec:prediction}

We will look at the out-of-sample predictive performance of three prediction methods based on FBRHT MCMC sampling results. Suppose we want to provide a predictive probability for the response $y_{*}$ of a test case with features $\bx_{*}$.  The first method is to average the predictive likelihood $P(y_{*} | \bx_{*}, \bbeta)$ with respect to $f(\bbeta|\by,\bX)$:
\begin{equation}\label{eqn:tprobit-avg}
P(y_{*} | \bx_{*}, \by, \bX) = \int P(y_{*} | \bx_{*}, \bbeta)f(\bbeta|\by,\bX) d\bbeta, 
\end{equation}
where $P(y_{*} | \bx_{*}, \bbeta))$ is given by the Robit model (eqn. \eqref{eqn:model-y}).  The above integral will be approximated by averaging $P(y_{*} | \bx_{*}, \bbeta)$ over MCMC samples of $\bbeta$.  We will refer to this prediction method by \textbf{\tavg}. \tavg~makes predictions by consolidating the information from all posterior modes (feature subsets) of $f(\bbeta|\by, \bX)$. In order to compare with other methods which report only a single feature subset, we are also interested in looking at the predictive power of a single feature subset found by FBRHT as described in Section~\ref{sec:mcmc-divide}. The two feature subsets that we will exam are the \textit{top} feature subset with the highest posterior frequency and the \textit{optimal} feature subset with the smallest cvAMLP. The prediction methods based on these two feature subsets are referred respectively as \textbf{\ttop}~ and \textbf{\topt}.  Suppose the feature subset is $\mS$ (top or optimal), the predictive probability is found with \texttt{bayesglm} as follows.   We apply \texttt{bayeglm} to find an estimate $\hat \bbeta$ from the dataset containing only the features in $\mS$, denoted by $(\by,\bX_{1:n,\mS})$; then we find the predictive probability with $P(y_{*}=c|\bx_{*,\mS}, \hat\bbeta)$, for $c = 0, 1$.  

The three prediction methods will be tested and compared to other methods in simulation studies and real data analysis.  In simulated studies, we average the predictive power measures (ER, AMLP, and AUC)  over multiple simulated datasets. In real data analysis, we find the predictive power measures by LOOCV. We will compare both the predictive performances of different methods and the number of features used by them.  To distinguish with the within-sample LOOCV predictive power measures of all feature subsets as described in Sec.~\ref{sec:fsubset-cv}, we will denote these out-of-sample predictive performance measures with ``ER'', `` AMLP'', and  ``AUC'' in the following empirical studies.

\section{Simulation Studies}\label{sec:sim}

\subsection{A Toy Example with Two Correlated Features} \label{sec:toy} \label{sec:tfs}

\begin{figure}[htp]
     \caption{Demonstration of within-group selection with two correlated features.} \label{fig:toy-mcmc}
     \centering
     \subfloat[][Scatterplot of Feature 1 and Feature 2\label{fig:tfs}]{
     \includegraphics[angle = 90, width=0.45\textwidth, height = 0.4\textwidth, trim = 1.4cm 0.7cm 1.9cm 0]{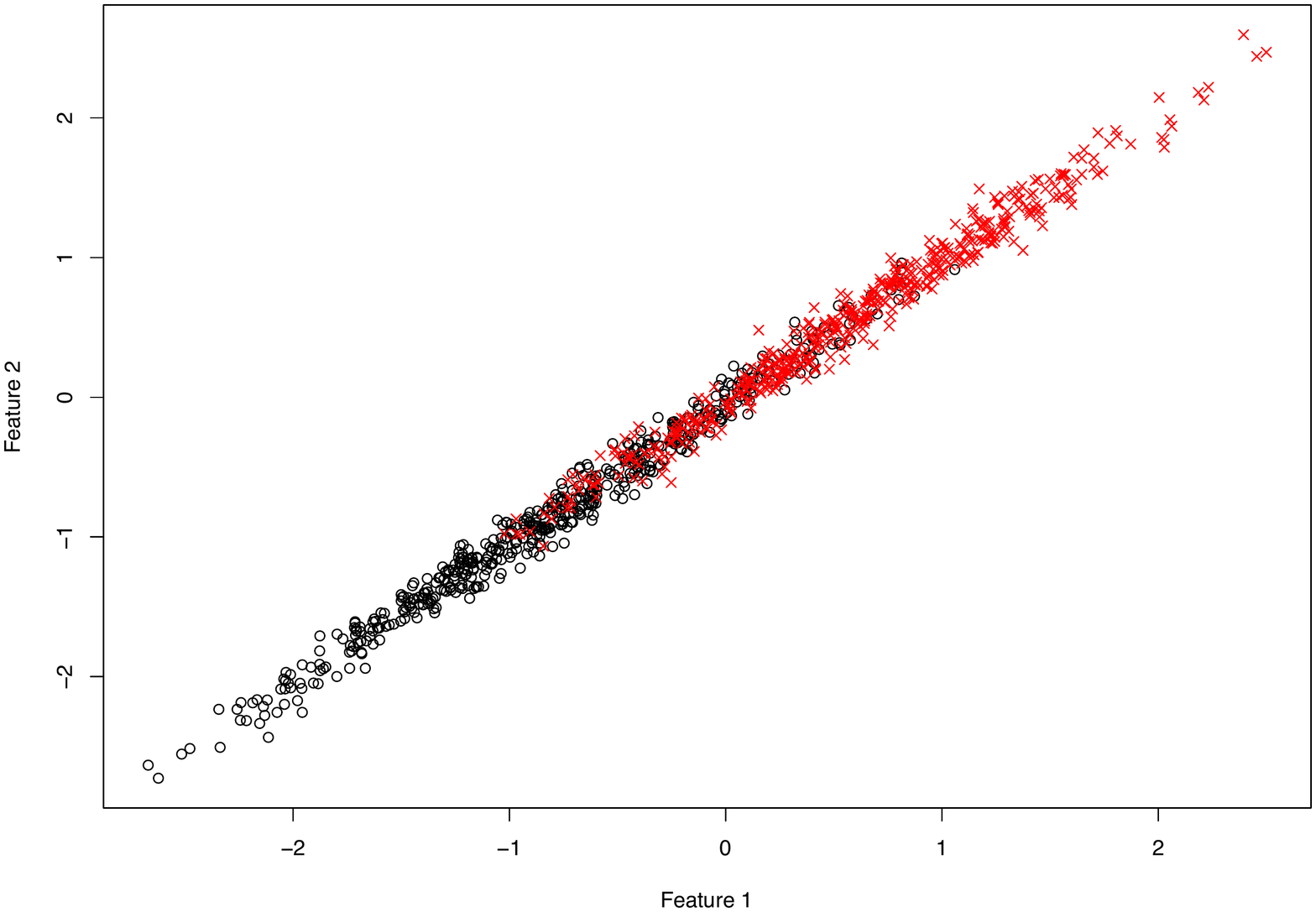}}
     \subfloat[][Scatterplot of MCMC samples of $(\beta_{1},\beta_{2})$\label{fig:tfs-mcmc}]{\includegraphics[width=0.45\textwidth, height = 0.4\textwidth]{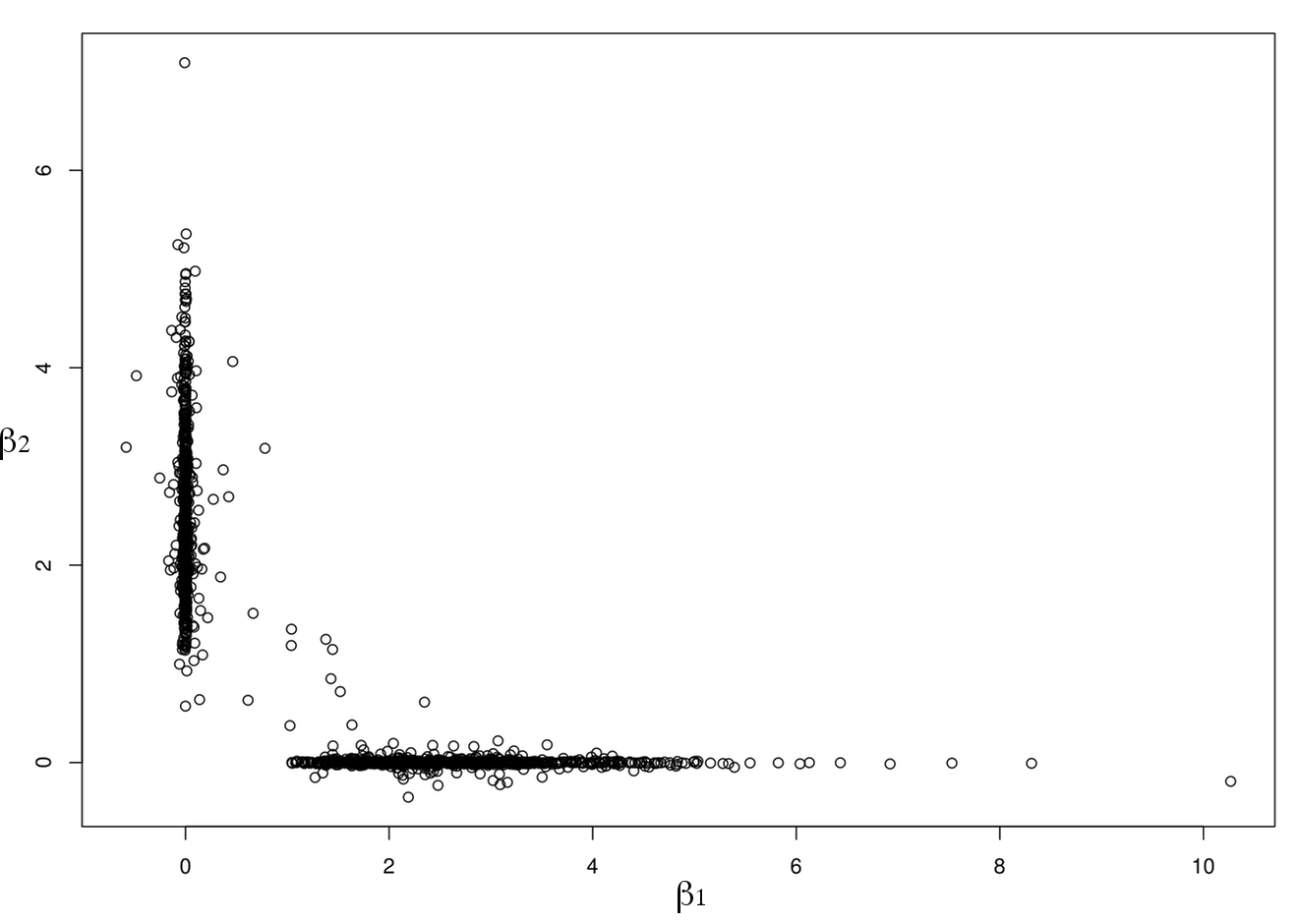}}
     \vspace*{-10pt}
\end{figure}

We first use a toy example to demonstrate the within-group selection property of  heavy-tailed $t$ priors with small scales. The response (class label) $y_{i}$ is equally likely to be $0$ or $1$, representing two classes. The means of the two features vary in two classes. The mean of feature $x_{j}$ in class $c$ is denoted by $\mu^{c}_{j}$, for $c=0,1$.  We fix $\mu^{0}_{j}=0$ and $\mu^{1}_{j}=2$ for $j=1,2$.   The response $y_{i}$ and feature values $\bx_{i}=(x_{i1}, x_{i2})$ for each case $i$ are generated as follows:
\begin{eqnarray}
&&P(y_i = c) = 1/2, \ \mbox{for } c=0,1 , \\
&&z_{i} \sim N(0,1), \ \epsilon_{ij} \sim N(0, 0.1^{2}),\ \mbox{for } j = 1,2\\
&&x_{ij} = \mu^{y_i}_{j}+ z_{i} + 0.1\epsilon_{ij},\ \mbox{for } j = 1,2.
\end{eqnarray}
Figure \ref{fig:tfs} shows the scatterplot of feature 1 and feature 2 in a simulated dataset after they are normalized to have mean 0 and standard deviation (sd) 1.  In the the above model,  feature 1 and feature 2 are both related to the class label $y_{i}$ because they have different means $\mu^{c}_{j}$ in two classes.  However, they are strongly correlated with a correlation coefficient 0.995  because they are generated with a common random variable $z_{i}$. Therefore, the two features together do not provide significantly more information than only one for predicting the class label $y_{i}$ (ie, separating the two classes). In real high-dimensional datasets, the size of such group of highly correlated features may be  very large. 

We generate a dataset of $n=1100$ cases using the above model. $100$ cases are used as training cases for fitting models, and the remaining $1000$ cases are used to test the predictive performance of fitted models. We fit the training dataset with FBRHT model by running the MCMC algorithm with $\alpha_{1}=1, w_{1}=\exp(-10)$ (i.e., Cauchy prior with scale $e^{-5}$) and other default settings specified in Sect.~\ref{sec:setting-mcmc} for 12000 iterations after a burn-in period. Figure \ref{fig:tfs-mcmc} shows the scatterplot of MCMC samples of coefficients  ($\beta_{1},\beta_{2}$).  These plots show that the posterior distribution of $(\beta_{1},\beta_{2})$ of FBRHT has two major modes, in each of which, only one coefficient has large non-zero value and the other one is shrunken to a value close to $0$. Each posterior mode represents a feature subset.

We divide the MCMC samples of $\bbeta$ into subpools according to their modes, and find subsets of selected features, using the method described in Sect.~\ref{sec:mcmc-divide}. For each subset, we also find its out-of-sample predictive power with  the test cases using the procedures described in Sect.~\ref{sec:prediction}. The feature selection and prediction results of FBRHT are shown in Table \ref{tab:tab_fss_tfs}a.  Two feature subsets with significant frequencies are found, each containing only one of feature 1 or 2.  The subset containing both of the features has very low frequency (0.02). The out-of-sample predictive power measures of each of the three subsets are very similar. Since the two features are highly correlated, the feature subset containing both of them does not give significantly better predictions, though a slightly better predictive performance is achieved by the subset using both of features; this is expected since this is the true model.  

\begin{table}[htp]
\caption{Feature subset selection and prediction results in a toy dataset with only two correlated features. The ``coefs'' of FBRHT are found by \bayesglm~with the data on the selected subset. An ``Inf'' AMLP  may happen as long as the predictive probability for a single case is extremely wrong.}\label{tab:tab_fss_tfs}

\centering
\begin{tabular}{rrrrrrr}
  
   \multicolumn{6}{l}{(a) Feature subsets selected by FBRHT} \\[5pt]
  
  fsubsets & freqs & coefs & AMLP & ER & AUC \\
  \hline
 1 & 0.56 & 2.62 &0.37 & 0.185 & 0.91 \\
 2 & 0.42 & 2.58 & 0.37 & 0.180 & 0.91 \\
1,2 & 0.02 &0.67, 1.94& 0.37 & 0.178 & 0.91 \\[5pt]

\multicolumn{6}{l}{(b) LASSO, PLR (\texttt{bayesglm}) and Random Forest (RF)} \\[5pt]

Method & & coefs & AMLP & ER & AUC \\
  \hline
LASSO & &1.15, 1.27 & 0.37 & 0.184 & 0.91 \\
RF& & 1.26, 1.26 & Inf & 0.219 & 0.88 \\
 PLR & &24.63, 24.53 & 0.37 & 0.184 & 0.91 \\
   \hline
\end{tabular}

\end{table}

The within-group selection provides a succinct feature subset to facilitate further investigation and comprehension for biologists. We clarify that  we do not lose other features because of the within-group selection, because we can still identify the other feature from the one selected in our feature subset using the correlation information among features if this is demanded in practice.

We also apply the LASSO, Penalized Logistic Regression (PLR) with hyper-LASSO penalty and Random Forest (RF with details given in Sect.~\ref{sec:others}) to estimate the coefficients and find predictive power of the estimated coefficients using the test cases. The results are shown in Table \ref{tab:tab_fss_tfs}b. We see that they all select both of the features, without making selection within the group. The prediction performances are not significantly better than the subset with only one feature found by FBRHT.  Note that the results shown by the LASSO are from a special case, because the LASSO can also make selection within group but not as aggressively as FBRHT does.  Generally, FBRHT with heavy-tailed priors with small scales select much fewer features than the LASSO from a group of highly correlated features. 


\subsection{An Example with Independent Groups of Features} \label{sec:indepgrp} \label{sec:ind}

In this section, we compare FBRHT  with other existing feature selection methods on simulated datasets with independent groups of features. Each dataset has $p = 2000$ features and $n = 1200$ cases, 200 of which are used as training cases and  the other 1000 cases are used as test cases. With $z_{ij}, \epsilon_{ij}, e_{i}$ generated from $N(0,1)$, we generate  the feature values $x_{ij}$ for $i=1,..., n,  j=1,...,p$ in four groups and the class label $y_{i}$ as follows:
\begin{eqnarray}
 x_{il}  =  z_{i1} + 0.5 \epsilon_{il}, i=1,...,n, l = 1,...,50 ,  \textbf{(Group 1)} &&
 \\
 x_{im}  =  z_{i2} + 0.5 \epsilon_{im}, i=1,...,n,  m = 51,...,100 , \textbf{(Group 2)}&&
 \\
 x_{ik}  =  z_{i3} + 0.5 \epsilon_{ik},i=1,...,n,    k = 101,...,150 , \textbf{(Group 3)}&&
 \\ 
 x_{ij}  \sim N(0,1), i=1,...,n,  j = 151,...,2000, \textbf{(Group 4)}&&
 \\
y_i   = 1 \mbox{ if } {(z_{i1} + z_{i2} + z_{i3})}/{\sqrt{3}} + 0.1 e_{i} > 0; = 0 \mbox{ otherwise.}&&
\end{eqnarray}
The $z_{i1}$, $z_{i2}$ and $z_{i3}$ are common factors for features in respective group. Since the features within each of Group 1-3 are related to a common factor, they are highly correlated. However, the features across groups are independent.  The response $y_{i}$ is generated with the values of the common factors $z_{i1}, z_{i2}, z_{i3}$. Therefore, $y_{i}$ is related to all the features in Group 1-3. The $y_{i}$ is unrelated to all the features in Group 4.  The true model of $y_{i}$ given $x_{ij}$ has non-zero coefficients for all features in Group 1-3. However, given a small number, $n$, of cases, we prefer to selecting only 1 feature from each correlated group for simpler interpretation.

We apply FBRHT and other methods including LASSO, Group LASSO (GL), supervised Group LASSO (SGL), Random Forest (RF), Penalized Logistic Regression (PLR) with hyper-LASSO penalty to fit the training cases and then test the predictive performance with the 1000 test cases.  The details of the implementation can be found from the appended Sections~\ref{sec:setting-mcmc} and~\ref{sec:others}. Particularly, we use the generic function \texttt{bayesglm} available in R package \texttt{arm} to fit logistic regression models penalized with Cauchy prior, whose scale is chosen with a default scheme \citep{gelman2008weakly}.  FBRHT is conducted with the default parameter settings as listed in Section~\ref{sec:setting-mcmc}.  As explained in Section~\ref{sec:mcmc-T-probit}, FBRHT is first run with all 2000 features in \textbf{stage 1}, and then rerun with $p^{*}=100$ top features selected with posterior means, both with the aforementioned settings. The feature selection and prediction use the MCMC samples in the \textbf{stage 2} with the top 100 features. Because of the large $p$ in stage 1, we ran FBRHT hours to ensure convergence. The stage 2 is fairly fast, usually a run with 5 mins can provide good results, however, we allowed FBRHT to run about 30 mins to obtain the results reported throughout this article.

We divide FBRHT MCMC samples based on a single dataset  into subpools using the method described in Section~\ref{sec:mcmc-divide}, and find a list of feature subsets. For each feature subset, we also find their cross-validatory predictive measures using the training cases as described in Section~\ref{sec:fsubset-cv}.  Table~\ref{tab:fsubset-indep} shows the top (by frequency) five feature subsets. According to the ``cvAMLP'', the top feature subset (1,57,140) is identified as the optimal feature subset too. We see that the top 4 feature subsets selected by FBRHT contain exactly one feature from each of  Group 1 - 3 (each with 50 features) and none from Group 4 (noise).

\begin{table}[htp]
\caption{Top 5 feature subsets selected by FBRHT, and their within-sample cross-validatory predictive power.  ``fsubsets'' gives I.D. of features in each subset, ``coefs'' is the vector of regression coefficients found with the posterior means, ``cvAMLP'' - ``cvAUC'' are cross-validatory predictive power measures of each feature subset, found with the method described in Section~\ref{sec:fsubset-cv}.}
\label{tab:fsubset-indep}
\centering
\begin{tabular}{rlrrrr}
  
    & fsubsets & freqs & cvAMLP & cvER & cvAUC \ \\
  \hline
    1 & 1,57,140 & 0.22 & 0.13 & 0.09 & 0.99 \ \\
    2 & 1,51,140 & 0.11 & 0.13 & 0.08 & 0.99 \ \\
    3 & 16,57,140 & 0.10 & 0.14 & 0.08 & 0.99 \ \\
    4 & 1,51,101 & 0.09 & 0.14 & 0.08 & 0.99 \ \\
    5 & 12,57 & 0.04 & 0.41 & 0.39 & 0.89 \ \\
   \hline
\end{tabular}
\end{table}

We test the out-of-sample predictive power of the optimal and top feature subset (which are the same in this example)  found from Table~\ref{tab:fsubset-indep} by applying \texttt{bayeglm} to make predictions for the 1000 test cases, and compare with other subsets containing only the top 3 features according to their absolute coefficient values. Table~\ref{tab:ind-top3} shows the comparison results. Clearly we see that the top and optimal feature subset selected by FBRHT has the best predictive power compared to all other feature subsets of the same size. This is because that other methods  fail to include one representative feature from each of Group 1 - 3 (signals) in the their top 3 features; they all miss one or more signal groups in their top 3 features. 

\begin{table}[htp]
\setlength{\tabcolsep}{3pt}
\caption{Comparison of out-of-sample predictive power of different subsets containing 3 features on a dataset with independent groups of features. The predictive measures are obtained by applying \texttt{bayesglm} to make predictions for the test cases.}
\label{tab:ind-top3}

\centering
\begin{tabular}{llrrr}
Method & fsubsets & AMLP & ER & AUC \\
  \hline
  \ttop & 1,57,140 & 0.22 & 0.10 & 0.97 \\ 
  \topt & 1,57,140 & 0.22 & 0.10 & 0.97 \\ 
  LASSO & 16,57,61 & 0.46 & 0.22 & 0.87 \\ 
  GL & 16,32,57 & 0.44 & 0.20 & 0.88 \\ 
  SGL & 16,138,140 & 0.47 & 0.24 & 0.86 \\ 
  RF& 28,50,67 & 0.46 & 0.22 & 0.86 \\ 
  PLR & 12,32,218 & 0.63 & 0.34 & 0.72 \\ 
 \hline
  \end{tabular}

\end{table}

We also compare the out-of-sample predictive power of the top and optimal feature subset found by FBRHT with the ``complete'' feature subsets selected by other methods.  For this comparison, we will make predictions for the test cases by directly using the estimated coefficients of respective methods.  We also include the predictive performance of the \tavg~\eqref{eqn:tprobit-avg} which makes predictions for the test cases by averaging over all MCMC samples, i.e., the ordinary Bayesian prediction.  Table~\ref{tab:prediction-1dataset-ind}b shows these predictive performance measures. We compare these predictive measures against the number of features used in making predictions, as shown in Table~\ref{tab:prediction-1dataset-ind}a. The numbers of features used in \ttop~and \topt~are just the number of features in the top and optimal subsets.  To count the number of features used by the methods other than FBRHT, we threshold their absolute coefficients by 0.1 of the maximum to decide whether or not they are used  in predictions. However, note that, the actual predictions by these methods use all the coefficients including those small ones. We choose 0.1 as a threshold because we use the same thresholding to obtain the top and and optimal feature subsets of FBRHT.   Furthermore, we divide the number of used features by their true group identities except for \tavg~due to complication.  Table~\ref{tab:prediction-1dataset-ind}a shows the number of selected features in each group after the thresholding.

\begin{table}[htp]
\setlength{\tabcolsep}{3pt}
\caption{Comparison of feature selection and out-of-sample prediction performance of different methods on a dataset with independent group of features. The number of features used by the others other than FBRHT are counted  after thresholding the absolute coefficients by 0.1 times the maximum. }\label{tab:prediction-1dataset-ind}

\centering


\begin{tabular}{lrrrrrrrrr}
\multicolumn{9}{l}{(a) Numbers of selected features in respective group}\\[3pt]  
    & \ttop & \topt & \tavg & LASSO & GL & SGL & RF & PLR \\
  \hline 
    Group 1 & 1 & 1  &-& 6 & 49 & 7 & 49 & 50 \\
    Group 2 & 1 & 1  &-& 5 & 50 & 10 & 49 & 50 \\
    Group 3 & 1 & 1  &-& 6 & 50 & 6 & 48 & 50 \\
    Group 4 & 0 & 0  &-& 13 & 341 & 12 & 14 & 1305 \\
    Total & 3 & 3  &$\leq$100& 30 & 490 & 35 & 160 & 1455 \\[5pt]
\multicolumn{9}{l}{(b) Out-of-sample predictive performance}\\[5pt]
    ER & 0.10 & 0.10 & 0.06 & 0.09 & 0.07 & 0.10 & 0.08 & 0.08 \\ 
  AMLP & 0.22 & 0.22 & 0.15 & 0.21 & 0.22 & 0.24 & 0.38 & 0.18   \\ 
  AUC & 0.97 & 0.97 & 0.99 & 0.97 & 0.99 & 0.97 & 0.98 & 0.98   \\
   \hline
\end{tabular}

\end{table}

   From Table~\ref{tab:prediction-1dataset-ind}, we see that the \tavg~has the best predictive performance even it uses no more than 100 features (in stage 2), which is better than the best performer (by ER and AUC) in non-FBRHT methods---Group LASSO with more than 490 features.  \ttop~and \topt~have slightly worse predictive performance than non-FBRHT methods, however, they use only 3 features, one from each signal group. In terms of efficiency in selecting useful features,  \ttop~and \topt~do the best jobs if we look at the ratio of predictive measure to number of used features.

We now look more closely at the feature selection results shown by Table~\ref{tab:prediction-1dataset-ind}a.   The top and optimal feature subset selected by FBRHT contains exactly 1 feature from each of Group 1-3 with signals, whereas other methods are all less sparse, selecting much larger subsets. Particularly, Group LASSO enforces the similarity of coefficients in each group, therefore, the estimated coefficients are all very small in magnitude and the differences among all coefficients are small. The consequence is that all the features in the signal groups along with a large number (341) of noise features are selected,  even we have used a fairly large threshold 0.1. The penalized logistic regression with Cauchy penalty (PLR, implemented with \bayesglm)  is expected to yield similar results as FBRHT. However, using the default scheme, \bayesglm~ chooses a much larger scale than the $e^{-5}$ used in FBRHT, and also gives homogeneous estimates of coefficients, similar to the estimates by Group LASSO, except that \bayesglm~ selects many more noise features than Group LASSO.  
If we chose very small scale as $e^{-5}$, on the other hand, \bayesglm~ will shrink all coefficients toward 0, resulting in very poor predictions. This example demonstrates that optimization algorithms have difficulty to find good modes from the posterior distribution based on non-convex penalty functions, though better algorithms could be developed by others. The MCMC algorithm we've developed can successfully travel across many such modes, finding sparse feature subset with very good predictive power. 

\begin{table}[htp]

\caption{Comparison of feature selection and out-of-sample prediction performance of different methods by averaging over 100 datasets with independent group of features. The number of features used by the others other than FBRHT in each dataset are counted  after thresholding the absolute coefficients by 0.1 times the maximum. }
\label{tab:prediction-100dataset-ind}

\centering
\setlength{\tabcolsep}{3pt}

\begin{tabular}{rrrrrrrrr}
\multicolumn{9}{l}{(a) Numbers of selected features in respective group}\\[5pt]  
 & \ttop & \topt & \tavg & LASSO & GL & SGL & RF & PLR  \\ 
  \hline
  Group 1 & 1.00 & 1.07 &-& 6.01 & 49.95 & 6.20 & 47.82 & 50.00  \\ 
  Group 2 & 1.00 & 1.08  &-& 6.00 & 49.94 & 5.99 & 47.48 & 50.00  \\ 
  Group 3 & 1.00 & 1.06  &-&5.94 & 49.95 & 6.04 & 48.34 & 50.00  \\ 
  Group 4 & 0.00 & 0.19  & -&14.44 & 401.33 & 8.83 & 3.78 & 1297.68  \\ 
 Total & 3.00 & 3.40 & $\leq$100& 32.39 & 551.17 & 27.06 & 147.42 & 1447.68  \\[5pt] 
\multicolumn{9}{l}{(b) Out-of-sample predictive performance}\\[5pt]
ER & 0.05 & 0.06 & 0.04 & 0.08 & 0.06 & 0.08 & 0.10 & 0.08  \\ 
  AMLP & 0.15 & 0.16 & 0.12 & 0.21 & 0.20 & 0.20 & 0.39 & 0.17  \\ 
  AUC & 0.99 & 0.99 & 0.99 & 0.98 & 0.99 & 0.98 & 0.97 & 0.98  \\ 
   \hline
\end{tabular}

\end{table}

We also replicated the studies on 100 different datasets. Table~\ref{tab:prediction-100dataset-ind} shows the averaged results of feature selection and predictive performance over 100 datasets. Similar conclusions can be drawn from Table~\ref{tab:prediction-100dataset-ind} as from Table~\ref{tab:prediction-1dataset-ind}.


\subsection{An Example with Correlated Weakly Differentiated Features} \label{sec:corrgrp}\label{sec:mul}

In this section we will compare the performance of FBRHT in 100 datasets generated such that two groups of features are weakly differentiated but have a strong joint effect on the response. Specifically, each dataset with $n = 1200$ cases and $p=2000$ features is generated as follows:
\begin{eqnarray}
 P(y_i=c) = \frac{1}{2}, \mbox{ for } c=1,2 ,&& \label{eqn:corgen1}
\\
x_{ij}  = \mu_{y_i,1} + z_{i1} + 0.5 \epsilon_{ij},\mbox{ for }    j=1,...,200, \textbf{(Group 1)} &&
 \\
x_{ij}  = \mu_{y_i,2} + 0.8 z_{i1} + 0.6 z_{i2} + 0.5 \epsilon_{ij},\mbox{ for }    j=201,...,400, \textbf{(Group 2)}&&
 \\
x_{ij}  = \mu_{y_i,3} + z_{i3} + 0.5 \epsilon_{ij}, \mbox{ for }  j=401,...,600, \textbf{(Group 3)}&&
 \\ 
x_{ij}  \sim N(0,1), \mbox{ for }  j=601,...,2000, \textbf{(Group 4)} && \label{eqn:corgen5}
\end{eqnarray}
where $z_{ij}$ and $\epsilon_{ij}$ are from $N(0,1)$, and the means of  features  in Group 1-3  in two classes are given by the following matrix $\mu_{c,1:3}$, where $\mu_{1,1:3} = (-0.3, 0.3, 1)$ and $\mu_{2,1:3} = (0.3, -0.3, -1)$. 
\begin{figure}[htp]
\caption{A Scatterplot of two weakly differentiated features with joint effect for $y_{i}$.}\label{fig:twocwd}
\centering
\includegraphics[width = 0.4\textwidth, height = 0.35\textwidth]{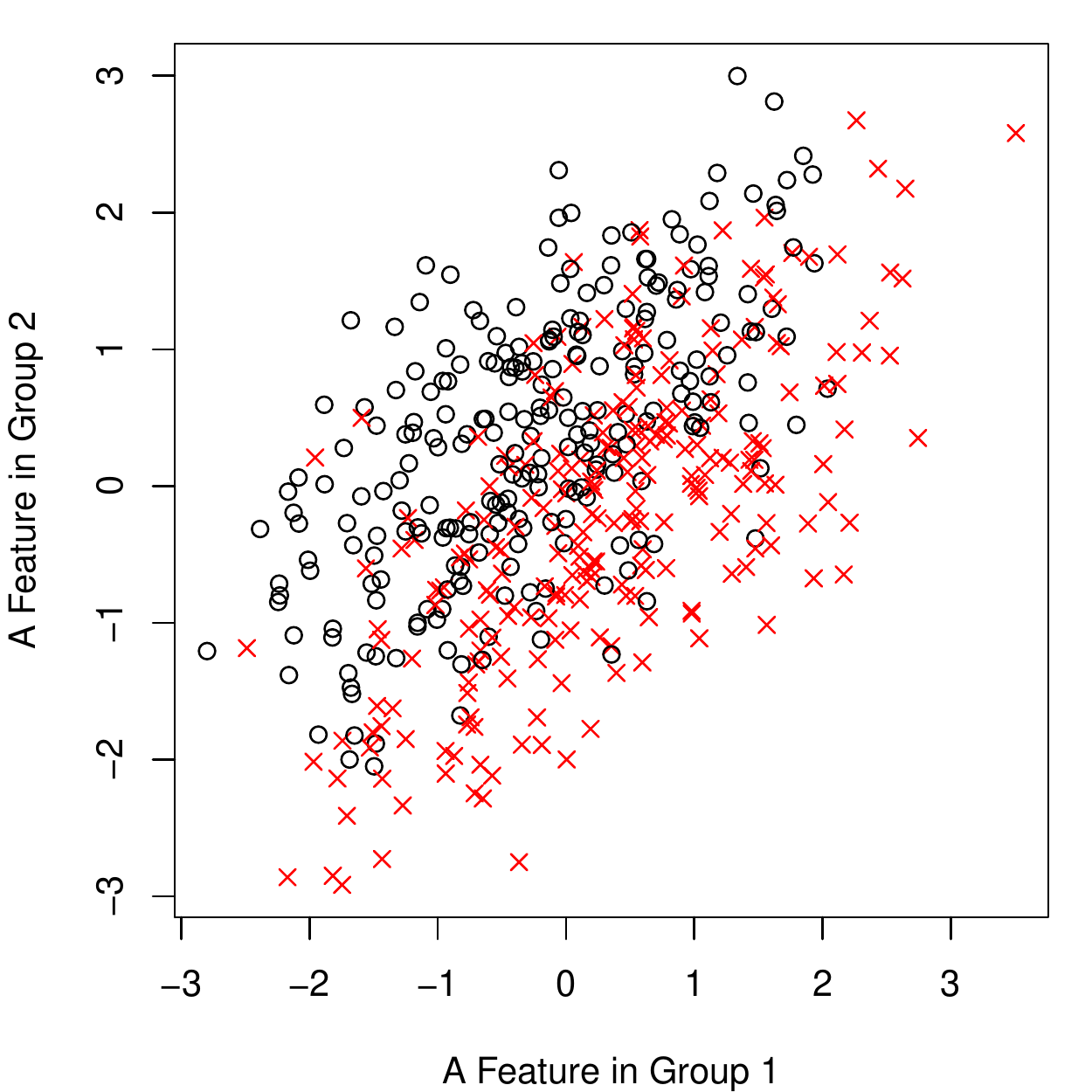}
\vspace*{-15pt}
\end{figure}

A dataset generated as above has 200 features in each of Group 1-3 related to the response and the remaining 1400 are completely noisy. Each feature in Group 1 and Group 2 is weakly differentiated due to the small difference in class means (0.3 vs -0.3).  The features within each group are positively correlated with correlation 0.8. Additionally, a feature from Group 1 and a feature from Group 2 has a correlation coefficient 0.64 because they share a common factor $z_{i1}$.  Therefore, a combination of two features from Group 1 and 2 respectively has clear joint effect for $y_{i}$, as shown by a scatterplot of two such features in Figure~\ref{fig:twocwd}.  

We run FBRHT and other methods to a dataset generated as above using the same procedures as used in Section~\ref{sec:indepgrp}. Table~\ref{tab:fsubset-cwd} shows the top 5 features subsets found by FBRHT, most of which are a subset containing only one feature from each of Group 1-3. 

\begin{table}[htp]
\caption{Top 5 feature subsets selected by FBRHT, and their within-sample cross-validatory predictive power. The table is read in the same way as Table~\ref{tab:fsubset-indep}. } \label{tab:fsubset-cwd}

\setlength{\tabcolsep}{3pt}
\centering
\begin{tabular}{rlrrrr}
     & fsubsets & freqs & cvAMLP & cvER & cvAUC \ \\
  \hline
    1 & 119,235,451 & 0.19 & 0.15 & 0.04 & 0.99 \ \\
    2 & 235,451 & 0.18 & 0.27 & 0.07 & 0.96 \ \\
    3 & 189,236,416 & 0.12 & 0.22 & 0.05 & 0.98 \ \\
    4 & 14,235,451 & 0.09 & 0.17 & 0.05 & 0.98 \ \\
    5 & 113,235,451 & 0.07 & 0.16 & 0.05 & 0.99 \ \\
   \hline
\end{tabular}
\end{table}

 We use 1000 test cases to compare the out-of-sample predictive power of the top and also optimal feature subset with other feature subsets containing only top 3 features found by respective methods. The results are shown by Table~\ref{tab:corgroup-top3}. We see that the top and optimal feature subset has better predictive power compared to the subsets found by other methods because they fail to include a representative feature from each of signal Group 1 - 3.  Particularly the feature subsets found by GL and PLR have very poor predictive performance. This is because that their coefficients are homogeneous, hence,  the features appearing in top 3 are chosen nearly randomly (indeed all noise), failing to represent the three signal groups. Other methods (LASSO, SGL, RF) select features only from one or two of the three signal groups in their top 3 features. 

\begin{table}[htp]
\setlength{\tabcolsep}{3pt}
\caption{Comparison of out-of-sample predictive power of different subsets containing 3 features. The table is read in the same way as Table~\ref{tab:ind-top3}.}
\label{tab:corgroup-top3}
\centering
\begin{tabular}{llrrr}
Method & fsubsets & AMLP & ER & AUC \\
  \hline
  \ttop & 119,235,451 & 0.41 & 0.17 & 0.91 \\ 
  \topt & 119,235,451 & 0.41 & 0.17 & 0.91 \\ 
  LASSO & 416,235,324 & 0.49 & 0.20 & 0.88 \\ 
  GL & 1532,1407,1461 & 0.77 & 0.49 & 0.51 \\ 
  SGL & 1532,324,14 & 0.61 & 0.27 & 0.79 \\ 
  RF & 587,595,527 & 0.41 & 0.18 & 0.90 \\ 
  PLR& 1532,1298,1407 & 0.78 & 0.49 & 0.52 \\ 
 \hline
  \end{tabular}
\end{table}

We also compare the predictive power of the top and optimal feature subsets found by FBRHT with ``complete'' feature subsets found by other methods using the same procedure for obtaining Table~\ref{tab:prediction-1dataset-ind}. We average the results over 100 datasets generated as above. The average results are shown in Table~\ref{tab:prediction-100dataset-cor}.  For these datasets, FBRHT methods have slightly worse predictive performance than Group LASSO and PLR, which successfully combine the power of all signal features from Group 1-3 to make better predictions. However, the feature subsets they select are the least sparse---containing 710.58 and 1564.26 features on average, including many noise features in Group 4. The feature subsets selected by other methods (LASSO, SGL, RF) are more sparse, but do not have significantly better predictive power than \ttop~and \topt~which uses only 2.63 and 3.81 features on average. In terms of the ratio of predictive measure to number of used features, \topt~is the best among all compared methods in this example.

\begin{table}[htp]

\caption{Comparison of feature selection and out-of-sample prediction performance of different methods by averaging over 100 datasets with correlated weakly differentiated features. This table is read in the same way as Table~\ref{tab:prediction-100dataset-ind}.}\label{tab:prediction-100dataset-cor}

\centering
\setlength{\tabcolsep}{3pt}

\begin{tabular}{rrrrrrrrr}
\multicolumn{9}{l}{(a) Numbers of selected features in respective group}\\[5pt]  
 & \ttop & \topt & \tavg & LASSO & GL & SGL & RF & PLR  \\ 
  \hline
Group 1 & 0.74 & 1.01 &- & 2.98 & 151.79 & 4.30 & 2.58 & 175.44  \\
  Group 2 & 0.81 & 1.09&-  & 3.16 & 149.41 & 5.29 & 4.63 & 177.12  \\
  Group 3 & 0.95 & 1.24 &- & 7.24 & 171.61 & 11.63 & 114.34 & 191.79 \\
  Group 4 & 0.13 & 0.47 &- & 9.40 & 237.77 & 21.61 & 2.85 & 1019.91  \\
 Total & 2.63 & 3.81 & $\leq$100& 22.78 & 710.58 & 42.83 & 124.40 & 1564.26  \\[5pt] 
\multicolumn{9}{l}{(b) Out-of-sample predictive performance}\\[5pt]
ER & 0.18 & 0.17 & 0.12 & 0.14 & 0.10 & 0.16 & 0.15 & 0.10  \\
  AMLP & 0.47 & 0.48 & 0.33 & 0.34 & 0.25 & 0.46 & 0.37 & 0.26  \\
  AUC & 0.90 & 0.91 & 0.95 & 0.93 & 0.97 & 0.92 & 0.93 & 0.97  \\
   \hline
\end{tabular}%
\end{table}

\section{Real Data Analysis}\label{sec:real}

\subsection{ Breast Cancer Methylation Data} \label{sec:breast}\label{section:breast}

Breast cancer is a type of cancer that begins in breast tissue. There are three different types of receptors on their surface and in their nucleus of breast cancer cells: human epidermal growth factor receptor 2 (HER2), estrogen receptor (ER) and progesterone receptor (PR). Cancer cells with ER (ER+) depend on estrogen to grow and so there are efficient treatments developed to block estrogen (e.g. tamoxifen) for controlling the growth of cancer cells.  In addition, for ER+ patients RT-PCR prognostic tests can be used to estimate the recurrence rate for suggesting possible choices of chemotherapy. However,  RT-PCR has limited predictive power for recurrence of cancer in ER- patients \citep{paik2004multigene}. Therefore, it is interesting to find diagnostic tools for accurately predicting ER status.  Looking for genetic biomarkers for classifying ER status is one promising approach. 

From Gene Expression Omnibus (GEO) repository, we downloaded a dataset studying ER status classification with genome-wide methylation profiling (GEO ID  \href{http://www.ncbi.nlm.nih.gov/geo/query/acc.cgi?acc=GSE31979}{GSE31979}), which were collected at Johns Hopkins Hospital and published by \citet{fackler2011genome-wide}. The dataset contains 48 samples with moderately estrogen receptor-positive (ER+), and 53 samples with receptor-negative (ER-).  The  DNA methylation level of each sample was measured with Human Methylation27 DNA Analysis BeadChip (GPL8490), which includes 27,578 probes. log2 transformation was applied to the original ratio of the methylation level.  

We used the method SAM (Significant Analysis of Microarrays by \citet{tusher2001significance}) to filter the original features (genes with methylation measurements),  and selected the top 5000 features for the following analysis based on classification models. We re-ordered the 5000 features according to the modified \tt-statistic of SAM. Therefore the  feature IDs given below are also ranks of their marginal correlations with ER status.   As we mentioned previously, univariate screening methods, such as SAM, may omit the weakly differentiated but perhaps useful features. However, note that the usefulness of weakly differentiated features also relies on their correlations with differentiated features as shown in Figure~\ref{fig:twocwd}.  The features with very low marginal correlations with the response are nearly constant across sample, hence, they have very correlations with differentiated features too. Therefore,  we use SAM to remove these features for reducing the computation time of fitting FBRHT.  


We first apply FBRHT to the whole dataset ($n=101$ samples, $p=5000$ features) with the same settings as used in the simulation studies. Table~\ref{tab:breasttop3}a shows the top 5 feature subsets and their LOOCV predictive  measures found with the whole dataset.  We see that FBRHT selects subsets containing only 2 or 3 features that have very good LOOCV predictive power. Particularly, we note that some features (eg. 366 and 1795) included in some subsets have very low marginal correlation rank, indicating that FBRHT selects them because of their joint effects for ER status. To compare FBRHT with other methods, we also find the LOOCV predictive measures of subsets of top 3 features selected by other methods; the results are shown in Table~\ref{tab:breasttop3}b.  We see that  the 3rd and 4th subsets found by FBRHT have significantly better predictive power for the ER+/- in the given dataset.  Finally, we point it out that since the dataset is used twice in both selecting these features and evaluating their predictive power, there is optimistic bias in these predictive measures, i.e., their predictive power may be worse in the whole population. In the above comparison, we assume that this bias is the same for all methods. 

\begin{table}[htp]
\caption{LOOCV predictive measures of feature subsets found from Breast Cancer Data. }
\label{tab:breasttop3}
\setlength{\tabcolsep}{2pt}
\centering
\resizebox{0.475\columnwidth}{!}{
\begin{tabular}{llrrrr}
    \multicolumn{5}{l}{(a) Feature subsets given by FBRHT} \\[5pt]
    & fsubsets & freqs & cvAMLP & cvER & cvAUC \\
  \hline
    1 & 23,77 & 0.05 & 0.21 & 9/101 & 0.98  \\
    2 & 77,554 & 0.03 & 0.25 & 11/101 & 0.96    \\
    3 & 1,366,1795 & 0.02 & 0.11 & 4/101 & 0.99  \\
    4 & 23,77,1587 & 0.02 & 0.16 & 6/101 & 0.99  \\
    5 & 1,1526 & 0.02 & 0.23 & 12/101 & 0.96  \\
   \hline
\end{tabular}
}   
\resizebox{0.5\columnwidth}{!}{ 
  \begin{tabular}{llrrr}
   \multicolumn{5}{l}{(b) Feature subsets given by other methods} \\[5pt]
    Method & fsubsets &  cvAMLP & cvER & cvAUC \\
    \hline
    LASSO & 25,266,614 & 0.27 & 10/101 & 0.95 \\
    GL & 2256,1795,266 & 0.52 & 21/101 & 0.82 \\
    SGL & 266,2256,1756 & 0.51 & 25/101 & 0.83 \\
    RF & 10,8,103 & 0.32 & 13/101 & 0.93\\
    PLR & 1,2256,4832 & 0.27 & 12/101 & 0.95 \\
    \hline
    \end{tabular}
} 
\end{table}

The feature subsets found by FBRHT may be of interest to biologists.  The detailed annotation information of these feature subsets provided by the original experiment platform and HGNC (HUGO Gene Nomenclature Committee) is given in appendix \ref{sec:anno}. The relationships between these genes and breast cancer ER status can be investigated further by biologists.



We also compare the out-of-sample predictive performance of FBRHT methods with the ``complete'' feature subsets found by the other methods as reported in Table~\ref{tab:prediction-100dataset-ind} for simulation studies. For real datasets, we instead use leave-one-out cross-validation.  We make  101 pairs of training and test datasets---in each pair, one of the 101 cases is left out as a test case, and the remaining 100 are used as training cases. In each pair of dataset, we use the training cases to select features  (for example finding the top and optimal feature subsets with FBRHT, and estimate the coefficients with the other methods), and build probabilistic classification rules, and then the classification rules are tested with the left-out case with the true class label. We average the predictive measures and also the number of used features over the 101 pairs of datasets.  For counting the number of used features, we also use 0.1 times the maximum to threshold their absolute coefficients for the methods we compared to as we use the same value to threshold MCMC samples of FBRHT for extracting feature subsets. Table~\ref{tab:prediction-breast} shows the comparison results.  From this table, we see that the predictive performance of \topt~is  slightly worse than LASSO and GL in terms of AMLP.  However, \topt~only uses  about 3 features on average, greatly smaller than the numbers of features used by all other methods. \ttop~is the best if we look at the ratio of predictive measure to number of used features. As we have seen in simulation studies, the feature selection results of Group LASSO and PLR are lack of sparsity, even though they have reasonable predictive performance.  For this dataset, \ttop~does not have good predictive performance as it tends to choose subsets of only 2 features. 

\begin{table}[htp]
\caption{Comparison of out-of-sample predictive performance on Breast Cancer Data.}
\label{tab:prediction-breast}
\setlength{\tabcolsep}{3pt}
\centering
\begin{tabular}{l*{8}r}
    &\topt&\ttop  & \tavg & LASSO & GL & SGL & RF & PLR \\
  \hline
    No. of Genes & 2.98 & 2.02 & $\leq 100$ & {39.57} & {2209.73} & {36.62 }& {187.63} & {2667.47} \\
    ER$\times$101  & 9 & 21 & {10} & 8 & 9 & 10 & 10 & 12 \\
    AMLP & 	0.33 & 0.51  & {0.33} & 0.28 & 0.27 & 0.42 & 0.34 & 0.33 \\
    AUC  &	0.96  & {0.88} & {0.91} & 0.94 & 0.94 & 0.95 & 0.93 & 0.94 \\
   \hline
\end{tabular}%
\end{table}

\subsection{Childhood Acute Leukemia Microarray Data} \label{sec:leuk}

In this section, we apply FBRHT and other methods to another microarray dataset related to two types of childhood leukemia: ALL (Acute lymphoblastic/lymphoid leukemia) and AML (Acute myeloid leukemia). From GEO repository, we downloaded the dataset with GEO ID \href{http://www.ncbi.nlm.nih.gov/geo/query/acc.cgi?acc=GSE7186}{GSE7186}, which was collected at Lund University Hospital and Linkoping University Hospital, and published by \citet{andersson2007microarray}.  The dataset contains 98 samples with ALL and 23 samples with AML. We preprocessed the data using the same procedure as in Section~\ref{section:breast} and kept $p=5000$ genes for the following analysis.

%
%
%
%
%
%

We apply FBRHT to the whole dataset ($n=121$ samples, $p=5000$ features) with the same settings as used previously. Table~\ref{tab:leuktop3}a shows a portion of feature subsets found by FBRHT and their LOOCV predictive measures.  Table~\ref{tab:leuktop3}b shows the subsets of top 2 features selected by the other compared methods. From the results shown by Table~\ref{tab:leuktop3}, we see that this dataset contains many genes with very good predictive power for AML and ALL.  The top 6 feature subsets  (some not shown in Table~\ref{tab:leuktop3}) found by FBRHT contains only a single feature that gives very good LOOCV predictive measures. In addition, FBRHT also identifies some subsets containing two features with slightly better predictive power,  such as  (32, 35), (30, 35), as shown in Table~\ref{tab:leuktop3}a. Most of the top 2 features identified by the other methods (except PLR) also have good predictive measures, but slightly worse than the subsets (32, 35) and (30, 35).  The top 2 features selected by PLR have significantly worse predictive power. We believe that this is due to the homogeneity in the coefficients of PLR. The results on PLR for this example are consistent with those from simulation studies.

\begin{table}[htp]
\caption{LOOCV predictive measures of feature subsets found from Acute Leukaemia Data. }
\label{tab:leuktop3}
\setlength{\tabcolsep}{2pt}
\centering
\resizebox{0.49\columnwidth}{!}{
\begin{tabular}{llrrrr}
    \multicolumn{5}{l}{(a) Feature subsets given by FBRHT} \\[5pt]
    & fsubsets & freqs & cvAMLP & cvER  & cvAUC \\
  \hline
    1 & 32 & 0.38 & 0.06 & 2/121 & 1.00 \\
    2 & 30 & 0.18 & 0.07 & 4/121 & 0.99 \\
    3 & 36 & 0.09 & 0.09 & 2/121 & 0.99 \\
    7 & 30,35 & 0.02 & 0.03 & 1/121 & 1.00\\
    8 & 32,35 & 0.02 & 0.03 & 1/121 & 1.00\\
   \hline
\end{tabular}
}   
\resizebox{0.475\columnwidth}{!}{ 
  \begin{tabular}{llrrr}
   \multicolumn{5}{l}{(b) Feature subsets given by other methods} \\[5pt]
    Method & fsubsets &  cvAMLP & cvER & cvAUC \\
    \hline
  LASSO & 32,35 & 0.03 & 1/121 & 1.00 \\
  GL & 35,115 & 0.15 & 4/121 & 0.95 \\
  SGL & 115,35 & 0.13 & 4/121 & 0.96 \\
  RF & 36,28 & 0.07 & 4/121 & 1.00 \\
  PLR & 1,5794 & 0.20 & 12/121 & 0.96 \\
    \hline
    \end{tabular}
} 
\end{table}

\begin{figure}[htp]
     \caption{Scatterplots of two feature subsets found from the leukamia data. } \label{fig:scatter_30_32}
     \centering
     \subfloat[][Gene subset (32, 35).]{\includegraphics[angle=0, width=0.45\textwidth, trim = 0cm 0cm 1cm 2cm]{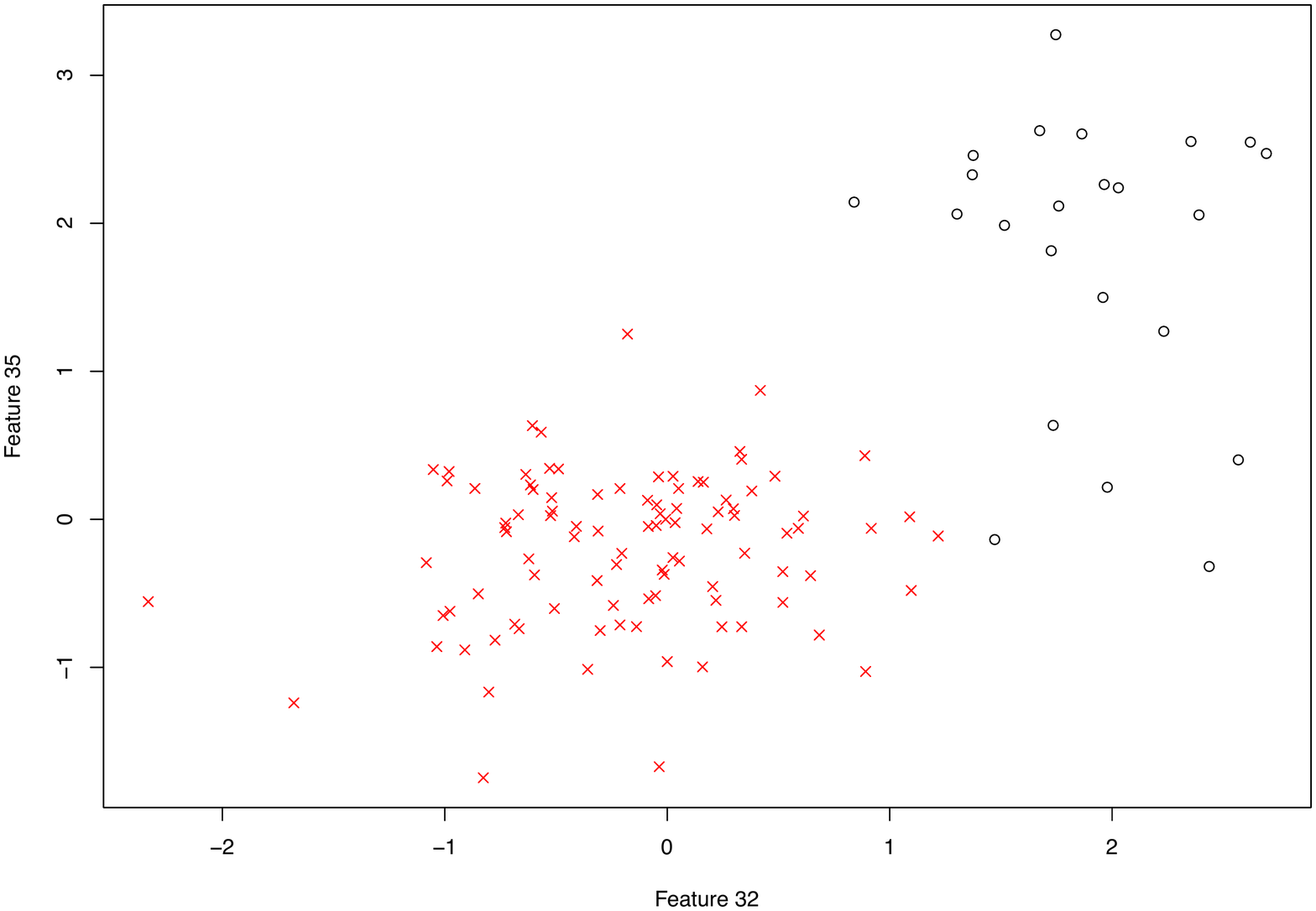}\label{fig:origin_leu8}} 
     \subfloat[][Gene subset (30, 35).]{\includegraphics[angle=90, width=0.45\textwidth, trim =  0cm  0cm 2cm 1cm]{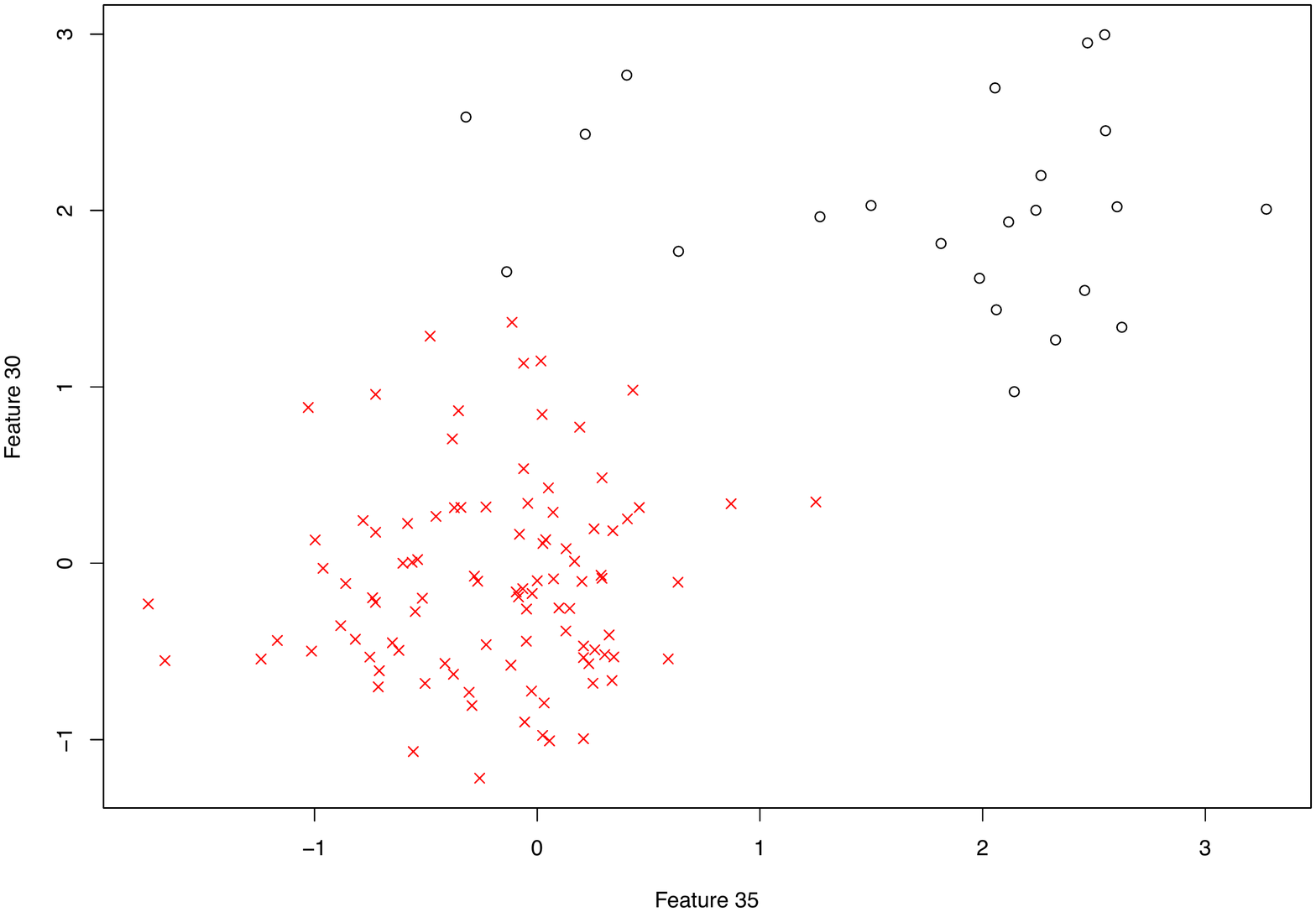}\label{fig:origin_leu7}}
     \vspace*{-25pt}
\end{figure}

To visualize the selected features from this dataset, we also draw the 2D scatterplots of two feature subsets, (32,35) and (30, 35), in Figure \ref{fig:scatter_30_32}.  We see that a combination of two features in each subset can form a very good separation of AML and ALL. From Figure~\ref{fig:origin_leu8} we also see that the gene 32 alone can nearly perfectly separate AML and ALL, which can also be seen from its LOOCV predictive measures given in Table~\ref{tab:leuktop3}a (the 1st entry). FBRHT successfully identifies gene 32 as the top feature subset and other succinct feature subsets as shown in Table~\ref{tab:leuktop3}a without including other redundant features.   

We also compare the out-of-sample predictive performances of FBRHT methods with the ``complete'' feature subsets selected by the other compared methods with LOOCV as described in Section~\ref{sec:breast}. The results are shown in Table~\ref{tab:prediction-leuk}. From what we have learnt from Table~\ref{tab:leuktop3} and Figure~\ref{fig:scatter_30_32}, a subset of one or two features can have very good power in separating ALL and AML in this dataset. However, using a quite large threshold 0.1 on relative coefficients, the other compared methods still select large numbers of features, with the most sparse subsets given LASSO which have 26.43 features on average. The lack of sparsity of the compared methods in feature selection is clearly demonstrated by this example.  \ttop~and \topt~ have slightly worse out-of-sample predictive measures than the other compared methods, however, they use only one or two features.  Finally, we want to point it out that the slightly larger ERs and AMLPs in \ttop~and \topt~are only caused by a few samples on the classification boundary.  Therefore, the AUC measures of the out-of-sample predictive probabilities given by \ttop~and \topt~are not much affected---still near the perfect value 1. 
\begin{table}[htp]
\caption[Comparison]{Comparison of out-of-sample predictive performance on Breast Cancer Data.}
\label{tab:prediction-leuk}
\setlength{\tabcolsep}{3pt}
\centering
\begin{tabular}{l*{8}r}
    & 	\ttop & \topt & \tavg &LASSO & GL & SGL & RF & PLR \\
  \hline
    No. of Genes & 1.00 &{1.95} & $\leq 100$ & {26.43} & {2783.26} & {50.34} & {149.33} & {3484.88} \\
    ER$\times 121$ & 	{3} &{5} & {2} & 1 & 0 & 2 & 2 & 10 \\
    AMLP & {0.07} & 	{0.09} &{0.09} & 0.04 & 0.05 & 0.03 & 0.12 & 0.34 \\
    AUC &{1.00} & {0.99} &{1.00} & 1.00 & 1.00 & 1.00 & 1.00 & 1.00 \\
   \hline
\end{tabular}%
\end{table}

\section{Conclusions and Discussions}\label{sec:remark}

In this paper, we have proposed a feature selection method, called fully Bayesian Robit with Heavy-tailed prior (FBRHT), which is based on the employment of MCMC to explore the posteriors of Robit classification models with heavy-tailed priors (hyper-LASSO penalization). Using simulation studies and real data analysis, we have demonstrated the success of FBRHT in finding sparse feature subsets with good predictive power.  Specifically, we have shown that FBRHT can efficiently discard noise features and also make selection within groups of correlated features automatically without a pre-specified  grouping structure.  Sparse feature subsets are much easier for understanding and  further more accurate experimental investigations.  Generally, we do not expect that the predictive power of a more sparse feature subset can be superior than that of a much larger subset.  Nevertheless,  our studies with simulated datasets and two real high-throughput datasets show that the predictive power of the sparse feature subsets selected by FBRHT are comparable with that of other much larger feature subsets selected by LASSO, group LASSO, supervised group LASSO, random forest, and logistic regression with non-convex penalization. In addition, our studies have shown that the sparse feature subsets selected by FBRHT have significantly better predictive power than the feature subsets of the same size taken from the top features selected by the aforementioned methods.

FBRHT reported here can still be improved in many aspects.  Our top priority is to improve the extraction of feature subsets from the MCMC samples. The method for dividing the MCMC samples reported in this paper still needs to be improved.  MCMC introduces small random jitters into a $\beta_{j}$ which should be very small in the mode. We use a fairly large threshold 0.1 on relative magnitudes of coefficients as an attempt to eliminate such jitters. However, we feel that this generally result in over-sparse feature subsets with risk of omitting features with small effects.   A better and more costly method is to find the exact modes using optimization methods. We need to come up with a fast optimization algorithm which can take the sparsity in coefficients into consideration. Other approaches such as mixture modelling and clustering can also be explored for dividing MCMC samples according to their modes.  Another approach is to find a feature subset from the MCMC samples that have the best matching (not the best within-sample predictive power) to the full MCMC samples using the so-called reference approach \citep{piironen2017comparison}.  A direct extension of FBRHT is to apply fully Bayesian inference to many other heavy-tailed priors as mentioned in the introduction section and other models such as linear models and graphical models. 

The MCMC sampling in stage 1 with large $p$ is fairly slow. However, the sampling in stage 2 with $p=100$ features is very fast.  The MCMC sampling in stage 1 with a large number of features may be unnecessary. From our empirical results, it seemly can be replaced by the fast LASSO, which does include representatives of from all signal groups in our simulation studies.The major problem of LASSO is the lack of sparsity. A follow-up MCMC exploration of the non-convex penalized likelihood as reported in this paper can be used to simplify the redundant feature subset found by the LASSO.

%% file: appendix.tex

\section*{Appendix}

\appendix

\section{Details of Computational Methods}
\label{sec:appendix}

\subsection{Setting Parameters of FBRHT } \label{sec:setting-mcmc}
We run MCMC sampling in two stages. In stage 1, we run MCMC sampling with all $p$ features. Then we use MCMC means of  $p$ coefficients to choose the top $p^{*}$ features. In stage 2, we run MCMC sampling with a reduced dataset with only the $p^{*}$ selected features from stage 1. Typically, we use the same MCMC sampling settings in two stages as listed below.  
\begin{itemize}

\setlength{\itemsep}{0pt}

\item Model specification parameters: $\alpha_{0}, \omega_{0}, \alpha_{1}, \omega_{1}$

The $\alpha_{0}$ and  $\sqrt{\omega_{0}}$ are the shape and scale parameter of $t$ distribution for modelling $y_{i}|\bxi$ in \eqref{eqn:model-y}.  The $\alpha_{1}$ and $\sqrt{\omega_{1}}$ are the shape and  scale parameter of $t$ distribution as the prior for $\beta_{j}$. They are all fixed at $\alpha_{0}=1, \omega_{0}=0.5, \alpha_{1}=1, \omega_{1}=\exp(-10)$ in most experiments if there is not specific mentioning. 

\item Restricted Gibbs sampling thresholding $\eta$

In step 3 of ``Restricted Gibbs sampling with HMC'' presented in Sec. \ref{sec:mcmc-T-probit}, we only choose $\beta_{j}$ with $j\in U = \{j|\hat\lambda_{j}>\eta\}$ to update with HMC.  We typically choose $\eta$ so that $10\%$ of $\bbeta$ are updated. 

\item HMC step size adjustment factor $\epsilon$ and lengths of trajectory $l_{1}$ and $l_{2}$

There are two critical tuning parameters for HMC:  the step size of each leapfrog step and the length of leapfrog trajectory. Fortunately they can be tuned independently \citep{neal2011mcmc}. Following \citet{neal2011mcmc}, we set leapfrog step size $\epsilon_{j}$ for $\beta_{j}$ with the second order derivative multiplied by a common adjustment factor $\epsilon$:
$
\epsilon_{j} = \epsilon \left( \frac{\partial^2 \mathcal U}{\partial \beta_{j}^2} \right)^{-1/2}.
$~
The $ \epsilon $ is an adjustment factor usually chosen from 0.1 to 1 such that we obtain the optimal rejection rate $30\%$ for HMC \citep{neal2011mcmc}. The required second-order derivative of  $\mathcal{U}$ with respect to $\beta_{j}$ is approximated by:
$
\frac{\partial^2 \mathcal U}{\partial \beta_j^2} \approx \sum\limits_{i=1}^n \frac{x_{ij}^2}{\hat\lambda_{j}} + {1\over \hat\lambda_{j}},
$~
where $x_{ij}$ are the value of the $j$th feature in the $i$th case.

The choice of length of trajectory is a little complicated.  \citet{neal1995bayesian} recommended to run HMC in two phases: initial (burn-in) phase and sampling phase. In initial phase, one uses a leapfrog trajectory of short length $l_{1}$ so that the log likelihood can be changed more quickly and the Markov chain can more quickly reach equilibrium or a local mode for our problems. In sampling phase, one should use a leapfrog trajectory of longer length $l_{2}$ to make full use of the ability of HMC to reach a distant point from the starting. $l_{2}$ is usually chosen after some pre-run experiments. Users may want to pre-run a Markov chain with a relatively large value of $l_{2}$ (e.g. 500,1000) and look at the trajectory of the magnitude of $\bbeta$. Because the leapfrog trajectory may go backwards to the starting point, $l_{2}$ should be chosen such that the magnitude of $\bbeta$ is explored in only one direction to the furthest extent without backtracking.  However, the optimal choice of $l_{2}$ is hard. It depends on specific problems. In addition, for our problems, the posterior are highly multi-modal, therefore, the optimal choice of $l_{2}$ may vary for different modes. An automatic scheme for choosing $l_{2}$, called NUTS,  is proposed by \citet{homan2014no-u-turn}. 

In our empirical studies, for the simplicity, we use $l_{2}= 50$ which appears sufficiently long for our problems.  We  set a shorter $l_{1} = 10$ in burn-in phase for faster convergence. 
 
\end{itemize}

\subsection{Implementation of Existing Feature Selection Methods}\label{sec:others}


\begin{itemize}\setlength{\itemsep}{0pt}

\item Penalized Logistic Regression using Hyper-LASSO penalty (PLR)

We use the function \bayesglm~ in the R package \texttt{arm} to fit Penalized Logistic Regression using Hyper-LASSO penalty.  The function \bayesglm~ uses the penalty based on  $ T(\alpha, \omega)$  prior,  the scaled $t$-distribution with shape parameter $\alpha$ and scale parameter $\sqrt{\omega}$. By default, \texttt{bayesglm} sets $\alpha_1=1$ and scale parameter $\sqrt{\omega_1}=2.5$ after the feature values are standardized in the suggested way \citep{gelman2008weakly}.

\item LASSO

LASSO is implemented using the R package \texttt{glmnet}. The choice of regularization parameter $\lambda$ is critical for the performance of LASSO. We feed the R function \texttt{glmnet} with a set of regularization parameters $\lambda = \{ \lambda_m, m= 1,2,...,M \}$. By default, we start with minimum $\lambda_1$ value $\lambda_1$ = 0.01 and choose $M$ = 100 candidate values with $\lambda_m = 0.01 m , m =1,2,...,M$. To find an optimal LASSO solution, we conduct cross-validation with respect to average minus log-probability over all candidate $\lambda_m$ values.  At last, we rerun \texttt{glmnet} on the whole dataset again with the optimal $\lambda$.  

\item Group LASSO  (GL)

We implement Group LASSO with prior group structure determined by hierarchical clustering (HC). We first conduct hierarchical clustering with the \texttt{hclust} function in the R package \texttt{clust} on the feature matrix $X$. For a given number of groups $C$,  the R function \texttt{hclust} can construct a tree with UPGMA (Unweighted Pair Group Method with Arithmetic Mean), and then the tree is cut into several groups by specifying the desired number of groups $C$.  The optimal value of $C$ is chosen using the maximum silhouette value from the set of $\{2, \ldots, 50\}$. With a chosen group structure (index), we can run Group LASSO (using the R function \texttt{gglasso}) on different values of the regularization parameter $\lambda$.  An optimal $\lambda$ is  chosen to minimize the cross-validated AMLP (average minus log-probability). At last we fit Group LASSO again with this optimal $\lambda$ and the given group structure. 

\item Supervised Group LASSO (SGL)

We use the same group structure as used for Group LASSO.  Given this group structure, SGL is implemented with a two-stage strategy. In stage 1,  for each feature group we then implement the LASSO algorithm with a reduced dataset and use the LASSO solution to extract significant features. More specifically, we fit LASSO (as we introduced before) with all the features in the $k$th group. The features with nonzero coefficients in the resulting LASSO solution will be retained and used as representatives of group $k$.  In stage 2,  all group representative features are then combined into a consolidated training dataset, with their group indices being retained. We then fit Group LASSO as described above on this consolidated dataset with the retained group indices. 

\item Random Forest (RF)

We implement Random Forest algorithm with the R package \texttt{RandomForest}  (based on Breiman and Cutler’s original Fortran code). Two important parameters in Random Forest are the number of trees (\texttt{ntree}) to grow and the number of variables randomly sampled as candidates at each split in the forest (\texttt{mtry}). With two arbitrary sets of candidate values for them, we fit randomForest with cross-validation. By default we use the candidate values of \texttt{mtry} ranging from $\sqrt{p}$ to $n$ if $\sqrt{p} < n$, or $n$ to $\sqrt{p}$ if $\sqrt{p} > n$. The candidate values of \texttt{ntree} are chosen from 250 to 500. For each pair value of \texttt{mtry} and \texttt{ntree} we run the Random Forest algorithm with the R function \texttt{randomForest} with cross-validation. The optimal pair values of \texttt{mtry} and \texttt{ntree} are then selected with respect to minimum AMLP. We then fit the whole dataset again with the optimal value of \texttt{mtry} and \texttt{ntree}. 
\end{itemize}

\subsection{An Investigation of Computational Efficiency of FBRHT}

In this section, we use a simulation experiment to briefly demonstrate the high efficiency of the sampler used in FBRHT. We will focus on the efficiency of FBRHT sampler in exploring multiple modes of Robit posterior distributions, by comparing to the JAGS \citep{plummer2003jags}, a black-box MCMC sampler. JAGS cannot scale well for very high-dimensional problems. Therefore, we simulate a dataset with only $p=100$ features for this comparison. Such examples also represent the stage-2 of FBRHT in which only a pre-selected small feature subset is used. However, we want to point it out that FBRHT works well in very high-dimensional problems such as with $p=5000$ in our real data analysis, except that the results are harder to interpret due to the large number of feature subsets.  We generate a dataset using a similar model described by equations~\eqref{eqn:corgen1}-\eqref{eqn:corgen5}, except that there are only 10 features in each of Group 1-3, and 70 features in Group 4. We run FBRHT and JAGS with Robit model for a long time (16835 seconds).  For FBRHT, we use the same settings as given in the appended Section~\ref{sec:setting-mcmc}.  We thin the original MCMC iterations into 1000 super-transitions in a way such that each super-transition (a transition consists of multiple original iterations) in FBRHT and JAGS costs the same time. We divide the 1000 MCMC samples produced by both FBRHT and JAGS to find feature subsets using the same way as described in Section~\ref{sec:mcmc-divide}.  We then monitor whether a mode switching occur from two consecutive super-transitions. Table~\ref{tab:cetime} shows the comparison results. Clearly, FBRHT sampler is much more efficient than JAGS in exploring a large number of modes of Robit posterior distributions. 

\begin{table}[htp]
\caption{Comparison of Computational Efficiencies of FBRHT and JAGS.}
\label{tab:cetime}
\centering
\begin{tabular}{ccc}
 &  Frequency of Mode Switching & No. of Found Modes \\
  \hline
JAGS  & 28/1000 & 9  \\
  FBRHT & 899/1000 & 70  \\
   \hline
\end{tabular}%
\vspace*{-10pt}
\end{table}

\bigskip
\section{Annotations of Selected Genes from Breast Cancer Methylation Data}\label{sec:anno}

\bigskip

\resizebox{0.98\columnwidth}{!}{%

\footnotesize

\begin{tabular}{ p{0.8cm}  p{1cm}  p{1.5cm} p{1.5cm} p{6.5cm}  p{2cm}}

   \hline
Feature ID & Gene ID & Gene Symbol & Synonym & Annotation & Gene  Product\\ 
\hline
1 & 100 & ADA & adenosine aminohydrolase &  Go function hydrolase activity, adenosine deaminase activity. go process nucleotide metabolism, purine ribonucleoside monophosphate biosynthesis, antimicrobial humoral response sensu Vertebrata. & adenosine deaminase\\ 
\hline
366 & 196472 & FAM71C & MGC39520 & synonym: MGC39520 & hypothetical protein LOC196472 \\ 
\hline
1795 & 25946 & ZNF385 & HZF; RZF; ZFP385; DKFZP586; G1122 & retinal zinc finger, go component: nucleus; go function: DNA binding; go function: zinc ion binding; go function: metal ion binding; go process: transcription; go process: regulation of transcription; DNA-dependent & zinc finger protein 385 \\ 
\hline
23 & 10164  & CHST4 & LSST & N-acetylglucosamine; 6-O-sulfotransferase; HEC-GLCNAC-6-ST; go component: membrane; go component: Golgi stack; go component: Golgi trans face; go component: integral to membrane; go component: intrinsic to Golgi membrane; go function: transferase activity; go function: N-acetylglucosamine 6-O-sulfotransferase activity; go process: inflammatory response; go process: sulfur metabolism; go process: carbohydrate metabolism; go process: N-acetylglucosamine metabolism  & carbohydrate (N-acetylglucosamine 6-O) sulfotransferase 4 \\ 
\hline
77 & 84816 & RTN4IP1 & NIMP; MGC12934  & NOGO-interacting mitochondrial protein; go function: zinc ion binding; go function: oxidoreductase activity & reticulon 4 interacting protein 1. \\ 
\hline
1587 & 9274 & BCL7C& BCL7C & B-cell CLL/lymphoma 7C & B-cell CLL/lymphoma 7C \\ 
\hline

\end{tabular}
}